\documentstyle[preprint,pra,aps,eqsecnum]{revtex}

\preprint{Submitted to Proceedings of the ITP Conference on Quantum %
Coherence and Decoherence}

\draft

\title{Information-theoretic approach to quantum error correction\\
and reversible measurement}

\author{
M.~A.~Nielsen,\thanks{%
Electronic address: mnielsen@tangelo.phys.unm.edu}$^{(1,2)}$
Carlton~M.~Caves,\thanks{%
Electronic address: caves@tangelo.phys.unm.edu}$^{(1,2)}$
Benjamin~Schumacher,$^{(3)}$ and
Howard Barnum$^{(1,2)}$ 
}

\address{$^{(1)}$Center for Advanced Studies, Department of Physics and 
Astronomy,\\ University of New Mexico, Albuquerque, NM 87131-1156}
\address{$^{(2)}$Institute for Theoretical Physics, University of California, 
Santa Barbara, CA 93106-4030}
\address{$^{(3)}$Department of Physics, Kenyon College, Gambier, OH 43022}

\date{\today}

\tightenlines

\begin{document}

\bibliographystyle{prsty}

\pagestyle{plain}
\pagenumbering{arabic}

\maketitle

\begin{abstract}
Quantum operations provide a general description of the state changes 
allowed by quantum mechanics.  The reversal of quantum operations is
important for quantum error-correcting codes, teleportation, and reversing
quantum measurements.  We derive information-theoretic conditions and
equivalent algebraic conditions that are necessary and sufficient for 
a general quantum operation to be reversible.  We analyze the thermodynamic 
cost of error correction and show that error correction can be regarded 
as a kind of ``Maxwell demon,'' for which there is an entropy cost 
associated with information obtained from measurements performed during 
error correction.  A prescription for thermodynamically efficient error 
correction is given.
\end{abstract}


\narrowtext

\section{Introduction}

Quantum operations arise naturally in the study of noisy quantum channels,
quantum computation, quantum cryptography, quantum measurements, and
quantum teleportation.  In each of these applications it is of interest 
to learn when a quantum operation can be {\em reversed}.  This paper gives 
a simple, physically meaningful set of necessary and sufficient conditions 
for determining when a general quantum operation can be reversed, thereby 
unifying and extending earlier work 
\cite{Schumacher96a,Schumacher96b,Nielsen96b}.  The picture we develop 
applies equally well to the reversal of quantum measurements, as in the 
processes described by Mabuchi and Zoller \cite{Mabuchi96a}, and to the 
protection of quantum states against decoherence, as described in the 
literature on quantum error correction (see, for example,~\cite{Calderbank96b} 
for references).  Finally, teleportation \cite{Bennett93a} can be understood 
as the reversal of a quantum operation, as was shown in \cite{Nielsen96b}, 
and the results obtained here are being applied in further work on 
characterizing schemes for teleportation.

The paper is organized as follows.  In Sec.~\ref{sect: qop} we review
the formalism of quantum operations and its application to the theory 
of generalized measurements, and we define the notion of a reversible 
operation.  Section~\ref{sect: measures} introduces information-theoretic 
measures associated with a quantum operation, which are used throughout 
the remainder of the paper.  These measures, {\em entanglement fidelity\/} 
and {\em entropy exchange\/}, were introduced earlier by one of us 
\cite{Schumacher96a} for the special case of deterministic operations; 
their definitions and properties are extended to general operations in 
Sec.~\ref{sect: measures}.  Section~\ref{sect: information characterization} 
states and proves an information-theoretic characterization of reversibility 
for general quantum operations, which is then given some simple applications 
in Sec.~\ref{sect: applications}.  Section~\ref{sect: algebraic reversal} 
presents an alternative algebraic description of when a quantum operation 
can be reversed.  Though many of the results in 
Sec.~\ref{sect: algebraic reversal} are already known, we provide new 
proofs, and the constructions used in these proofs are important in 
Sec.~\ref{sect: second law}, where we give a thermodynamic analysis of
error correction and show that schemes for performing perfect error 
correction can be done in a thermodynamically efficient way.  
Section~\ref{sect: conc} offers concluding remarks.

\section{Quantum operations}
\label{sect: qop}

\subsection{Definition and characterizations}

A simple example of a state change in quantum mechanics is
the unitary evolution experienced by a closed quantum system. 
The final state of the system is related to the initial
state by a unitary transformation $U$,
\begin{equation} \label{eqtn: unitary operation}
\rho \rightarrow {\cal E}(\rho) = U \rho U^{\dagger}\;. \end{equation}
Unitary evolution is, however, not the most general type of state 
change possible in quantum mechanics. Other state changes, not 
describable by unitary transformations, arise when a quantum system 
is coupled to an environment or when a measurement is performed on 
the system.

How does one describe the most general possible quantum-mechanical
dynamics that takes input states to output states?  The answer to this 
question is provided by the formalism of ``quantum operations.''  This 
formalism is described in detail by Kraus \cite{Kraus83a} and is given 
a short, but quite informative review in an appendix to~\cite{Schumacher96a}. 
In this formalism the input state is connected to the output state by the 
state change 
\begin{equation} \label{eqtn: gen change}
\rho \rightarrow 
\frac{{\cal E}(\rho)}{\mbox{tr}\bigl({\cal E}(\rho)\bigr)}\;,
\end{equation}
which is determined by a {\em quantum operation\/} ${\cal E}$.  The
quantum operation is a linear, trace-decreasing, completely positive map.
{\em Trace decreasing\/} means that $\mbox{tr}\bigl({\cal E}(\rho)\bigr)\le1$ 
for all normalized density operators $\rho$.  {\it Complete positivity\/}
means that in addition to preserving the positivity of density operators, 
the map preserves the positivity of all purifications of density operators.  
The trace in the denominator is included in order to maintain the 
normalization condition $\mbox{tr}(\rho)=1$.  

The most general form for a completely positive map ${\cal E}$ can be 
shown to be \cite{Kraus83a,Hellwig70}
\begin{equation} \label{eqtn: gen op}
{\cal E}(\rho) = \sum_j A_j \rho A_j^{\dagger}\;. 
\end{equation}
The system operators $A_j$, which must satisfy 
\begin{equation}
\sum_j A_j^\dagger A_j\le I
\end{equation}
in order that ${\cal E}$ be trace decreasing, completely specify the 
quantum operation.  We call Eq.~(\ref{eqtn: gen op}) an 
{\em operator-sum decomposition\/} of the operation, and we refer to 
the operators $A_j$ as {\em decomposition operators}.  

The operator-sum decomposition for a quantum operation is not unique, 
in that another set of decomposition operators $\{ A_j\}$ can give 
rise to the same operation.  For example, the operation on a 
spin-$\frac 12$ system defined by 
\begin{equation}
{\cal E}(\rho) = \frac{I}{\sqrt 2}\,\rho\frac{I}{\sqrt 2} +
	\frac{\sigma_z}{\sqrt 2}\,\rho\frac{\sigma_z}{\sqrt 2}
\end{equation}
can also be written in the form
\begin{equation}
{\cal E}(\rho) = \frac{I+\sigma_z}{2}\,\rho\frac{I+\sigma_z}{2} +
	\frac{I-\sigma_z}{2}\,\rho\frac{I-\sigma_z}{2}\;.
\end{equation}

Choi \cite{Choi75a} has classified all sets of decomposition operators 
that give rise to the same operation.  The result is that two sets of 
decomposition operators, $\{ A_k \}$ and $\{ B_j \}$, give rise to the 
same quantum operation if and only if they are related linearly by 
a square unitary matrix $u$: 
\begin{equation} \label{eqtn: decompositions}
B_j = \sum_k u_{jk} A_k\;.
\end{equation}
It is generally necessary to add some zero decomposition operators
to the set with the smaller number of elements so that both sets have
the same number of decomposition operators.  We call a decomposition
${\em minimal}$ if no decomposition into a smaller number of operators
exists; a decomposition is minimal if and only if the operators in the 
decomposition are linearly independent.

We say that an operation ${\cal E}$ is {\em pure\/} if it can be
written in terms of an operator-sum decomposition that contains only
one operator; that is, there exists an operator $A$ such that
\begin{equation}
{\cal E}(\rho)=A\rho A^\dagger\;.
\label{eqtn: pureoperation}
\end{equation} 
The unitary transformation~(\ref{eqtn: unitary operation}) is an 
example of a pure quantum operation.

We say that an operation ${\cal D}$ is {\em deterministic\/} or 
{\em trace-pre\-serving\/} if $\mbox{tr}\bigl({\cal D}(\rho)\bigr)=1$ 
whenever the input is a normalized density operator $\rho$.  For a 
deterministic operation, the decomposition operators $D_j$ satisfy a 
completeness relation 
\begin{equation}
\sum_j D_j^{\dagger} D_j = I\;,
\end{equation}
which implies that $\mbox{tr}\bigl({\cal D}(\rho)\bigr)=1$. Notice
that a pure deterministic operation must be a unitary transformation.
Any deterministic quantum operation ${\cal D}$ can be obtained by 
adjoining an ancilla system to the system of interest, allowing the 
system plus ancilla to interact unitarily, and then discarding the 
ancilla.  Such a dynamics leads to a state change of the form
\begin{equation} \label{eqtn: detop}
\rho \rightarrow 
\mbox{tr}_A\bigl ( V (\rho \otimes \sigma^A) V^{\dagger} \bigr )
\equiv {\cal D}(\rho)\;,
\end{equation}
where $\mbox{tr}_A$ denotes tracing out the ancilla, $\sigma^A$ is the 
initial state of the ancilla, and $V$ is the unitary operator for the
joint dynamics of the system and ancilla.  

{}For a general quantum operation, 
\begin{equation}
\mbox{tr}\bigl({\cal E}(\rho)\bigr)=
\mbox{tr}\Biggl(\rho\sum_j A_j^{\dagger}A_j\Biggr)
\end{equation}
is generally less than one, so ${\cal E}(\rho)$ must be renormalized,
as in Eq.~(\ref{eqtn: gen change}), to produce an output density operator.
A general quantum operation cannot be represented solely in terms of the 
joint unitary dynamics of the system and an ancilla, for that always leads
to a deterministic operation, as in Eq.~(\ref{eqtn: detop}).  A general
quantum operation can be obtained, however, if the joint dynamics is
followed by a measurement on the ancilla; the quantum operation corresponds
to particular measurement outcome described by an ancilla projection 
operator $P^A$.  The resulting state change, once the ancilla is discarded, 
is given by \cite{Kraus83a,Hellwig70}
\begin{equation} \label{eqtn: measmodel}
\rho \rightarrow 
{
\mbox{tr}_A\bigl((I \otimes P^A) V (\rho \otimes \sigma^A) V^{\dagger} 
(I\otimes P^A) \bigr)
\over 
\mbox{tr}\bigl((I \otimes P^A) V (\rho \otimes \sigma^A) V^{\dagger}
(I\otimes P^A) \bigr) 
} \equiv 
{{\cal E}(\rho)\over\mbox{tr}\bigl({\cal E}(\rho)\bigr)}\;.
\end{equation}
Notice that $\mbox{tr}\bigl({\cal E(\rho)}\bigr)$ is the probability of 
the measurement result described by $P^A$.  A deterministic operation 
arises in the special case $P^A = I^A$. We note that for any quantum
operation, there is a representation of the form~(\ref{eqtn: measmodel}) 
in which the initial ancilla state $\sigma^A$ is a pure state.

Suppose the measurement on the ancilla is described by a complete set
of orthogonal projection operators $P^A_i$, where the index $i$ labels 
the measurement outcomes.  Outcome $i$ corresponds to a quantum operation
\begin{equation}
{\cal E}_i(\rho)=
\mbox{tr}_A\bigl((I \otimes P^A_i)V(\rho\otimes\sigma^A)V^{\dagger}
(I\otimes P^A_i)\bigr) 
=\mbox{tr}_A\bigl((I \otimes P^A_i)V(\rho\otimes\sigma^A)V^{\dagger}\bigr)\;,
\label{eqtn: opi}
\end{equation}
which gives the unnormalized post-measurement state of the system, 
conditioned on outcome $i$.  The probability for result~$i$ is
\begin{equation}
p_i = \mbox{tr}\bigl({\cal E}_i(\rho)\bigr) \;.
\end{equation} 
If one discards the measurement outcome, the output density operator is
obtained by averaging over the outcomes, and the state change is given
by a deterministic quantum operation:
\begin{equation}
\rho \rightarrow
\sum_i p_i
{{\cal E}_i(\rho)\over\mbox{tr}\bigl({\cal E}_i(\rho)\bigr)}=
\sum_i {\cal E}_i(\rho)={\cal D}(\rho)\;.
\label{eqtn: measurement sum}
\end{equation}
Thus a deterministic quantum operation can always be regarded as
describing a measurement on the ancilla, whose result is discarded.

In the case of a deterministic operation, the ancilla can be regarded
as the system's environment; interaction with the environment gives
rise to the nonunitary evolution that is described by the operation.  For 
a general operation, the ancilla can also be regarded as an environment 
that can be observed and thus that has features of a measuring apparatus.  
Keeping these connotations in mind, we use the terms ancilla and 
environment interchangeably in the remainder of the paper.

\subsection{Operations and generalized measurements}

The connection of quantum operations to quantum measurements is easy
to explain.  Standard textbook treatments describe quantum measurements in
terms of a complete set of orthogonal projection operators for the system 
being measured.  This formalism, however, does not describe many of the 
measurements that can be performed on a quantum system. The most general 
type of measurement that can be performed on a quantum system is known 
as a {\em generalized measurement\/} \cite{Kraus83a,Gardiner91a,Peres93a}. 

Generalized measurements can be understood within the framework of 
quantum operations, because any generalized measurement can be performed 
by allowing the system to interact with an ancilla and then doing a
standard measurement described by orthogonal projection operators on 
the ancilla.  Thus, as we can see from Eq.~(\ref{eqtn: opi}), the most
general type of quantum measurement is described by a set of quantum
operations ${\cal E}_i$, where the index $i$ labels the possible 
measurement outcomes.  The sum of the operations for the various
outcomes is required to be a deterministic quantum operation, as in
Eq.~(\ref{eqtn: measurement sum}).  

Since we can give an operator-sum decomposition for each operation,
\begin{equation}
{\cal E}_i(\rho) =
\sum_j A_{ij} \rho A_{ij}^{\dagger}\;, \end{equation}
we can also say that the generalized measurement is completely described
by the system operators $A_{ij}$, which are labeled by two indices, $i$ 
and $j$, and which satisfy the completeness relation
\begin{equation} \label{eqtn: completeness}
\sum_{i,j} A_{ij}^{\dagger} A_{ij} = I\;. 
\end{equation}
If result $i$ occurs, the unnormalized state of the system immediately 
after the measurement is given by
\begin{equation}
{\cal E}_i(\rho) =
\sum_j A_{ij} \rho A_{ij}^{\dagger}\;. 
\end{equation}
The probability for result $i$ to occur is
\begin{equation}
p_i = \mbox{tr}\bigl({\cal E}_i(\rho)\bigr) =
    \mbox{tr}\Biggl(\rho\sum_j A_{ij}^{\dagger}A_{ij}\Biggr)\;. 
\end{equation}
This form makes the connection to the formalism of positive-operator-valued 
measures.  The operators 
\begin{equation}
E_i \equiv  \sum_j A_{ij}^\dagger A_{ij}
\end{equation}
are elements of a decomposition of the unit operator into positive
operators, as in the completeness relation~(\ref{eqtn: completeness}).
Such a decomposition of unity is called a {\em positive-operator-valued
measure\/} (POVM).

We say a measurement is {\em pure\/} if for each measurement result 
$i$, the corresponding quantum operation ${\cal E}_i$ is pure; that is, 
there exist operators $A_i$ such that
\begin{equation} \label{ideal state change}
{\cal E}_i(\rho) = A_i \rho A_i^{\dagger}\;. \end{equation}
The probability that result $i$ occurs is given by
\begin{equation}
p_i = \mbox{tr}(\rho A_i^{\dagger}A_i) \;,\end{equation}
It can be shown that pure measurements correspond to extracting the
maximum amount of information about the system from the state of the
apparatus to which the system is coupled.  

\subsection{Reversal of a quantum operation}
\label{sect: reversal}

When we talk about reversing a quantum operation ${\cal E}$, we generally 
do not mean that ${\cal E}$ can be reversed for all input states, but
rather only that ${\cal E}$ can be reversed for all input density operators 
$\rho$ whose support lies in a subspace $M$ of the total state space $L$. 
In the case of a trace-preserving operation ${\cal E}$, the subspace $M$ 
is sometimes called a {\em quantum error-correcting code\/} or simply a 
{\em code}.  It makes sense to talk about reversing ${\cal E}$ on a subspace 
$M$ only if ${\cal E}(\rho)\ne0$ for all $\rho$ whose support lies in $M$, 
and we assume this condition henceforth.  We say that a quantum operation 
${\cal E}$ is {\em reversible\/} on a subspace $M$ if there exists a 
{\em deterministic\/} quantum operation ${\cal R}$, acting on the total 
state space $L$, such that for all $\rho$ whose support lies in $M$,  
\begin{equation}
\rho = {\cal R}\!\left(
\frac{{\cal E}(\rho)}{\mbox{tr}\bigl({\cal E}(\rho)\bigr)} \right)=
\frac{{\cal R}\circ{\cal E}(\rho)}
{\mbox{tr}\bigl({\cal E}(\rho)\bigr)}\;.
\label{eqtn: reversalcondition}
\end{equation}
Here ${\cal R}\circ{\cal E}$ denotes the composition of ${\cal R}$ with 
${\cal E}$, that is, 
${\cal R}\circ{\cal E}(\rho)\equiv{\cal R}\bigl({\cal E}(\rho)\bigr)$.
We require the reversal operation ${\cal R}$ to be deterministic because 
we want the reversal definitely to occur, not just to occur 
with some probability, conditional on some measurement result or ancilla 
state.

In~\cite{Schumacher96b} the problem of reversing deterministic
quantum operations was considered. This case is of particular interest 
in situations where one is unable to obtain information about the 
environment.  In contrast, \cite{Nielsen96b} and~\cite{Mabuchi96a} 
considered reversal of operations representing measurements, in which 
case information about the environment is available.

We say that a measurement is {\em reversible\/} on a subspace 
$M$ of the total state space $L$ if for each measurement result $i$, 
the corresponding quantum operation is reversible.  Outcomes that have 
zero probability on $M$ are irrelevant, because they never occur, and 
thus they can be discarded.  We could define measurements that are only 
sometimes reversible by requiring that only some of the measurement 
results have reversible quantum operations.  Although we do not deal 
explicitly with such sometimes reversible measurements in this paper, 
the results obtained in Secs.~\ref{sect: information characterization}
and \ref{sect: algebraic theorem}, since they are derived for individual 
quantum operations, apply to sometimes reversible measurements.

\section{Information-theoretic measures for quantum operations} 
\label{sect: measures}

Schumacher \cite{Schumacher96a} introduced entanglement fidelity and
entropy exchange as useful in\-for\-ma\-tion-theoretic measures for 
characterizing deterministic quantum operations.  This section extends 
to general quantum operations the definitions of entanglement fidelity
and entropy exchange and generalizes the properties of those quantities
obtained in~\cite{Schumacher96a} and in~\cite{Schumacher96b}.   We begin 
by outlining the particular method for characterizing quantum operations 
that was used in~\cite{Schumacher96a} and~\cite{Schumacher96b} and that we 
use throughout the remainder of this paper.

\subsection{Method for characterizing quantum operations}
\label{sect: operation chacterization}

Suppose we have a quantum system, denoted henceforth by $Q$, and a 
quantum operation ${\cal E}$ that acts on states of $Q$.  We denote 
the dimension of the Hilbert space of $Q$ by $D$.  It is convenient 
to introduce two mathematical artifices, a reference system $R$, whose
Hilbert space has the same dimension, $D$, as the Hilbert space of $Q$, 
and an environment $E$, which has a Hilbert space of arbitrary dimension.

The joint state of the system $Q$ and the reference system $R$ is chosen 
so as to {\em purify\/} the initial state of $Q$; that is, $RQ$ is 
initially in a pure state 
$\rho^{RQ}=\bigl|\Psi^{RQ}\bigr\rangle\bigl\langle\Psi^{RQ}\bigr|$ 
satisfying
\begin{equation}
\mbox{tr}_R\Bigl(\bigl|\Psi^{RQ}\bigr\rangle
\bigr\langle\Psi^{RQ}\bigr|\Bigr) = \rho^Q\;,
\end{equation}
where $\rho^Q$ is the initial state of system $Q$. To reduce the clutter
in the notation, we drop the $Q$ superscript when it is clear that we are 
dealing with the primary quantum system $Q$.  The initial state of the 
environment $E$ is assumed to be a pure state $\rho^E = |e\rangle \langle e|$, 
which is uncorrelated with the system $RQ$.  Thus the initial state of the 
overall system is also pure:
\begin{equation}
\bigl|\Psi^{RQE}\bigr\rangle = \bigl|\Psi^{RQ}\bigr\rangle 
\otimes |e\rangle .
\end{equation}

The joint system $QE$ is subjected to a two-part dynamics consisting 
of a unitary operation, $U^{QE}$, followed by a projection onto the 
environment alone, described by a projector $P^E$.  The reference system
$R$ has no internal dynamics and does not interact with $Q$ or $E$.  
This two-part dynamics leaves the overall state pure.  As we mentioned
above, it is always possible to find $|e\rangle$, $U^{QE}$, and $P^E$ 
such that
\begin{equation}
{\cal E}(\rho^Q)= 
\mbox{tr}_E
\bigl(P^E U^{QE}(\rho^Q\otimes\rho^E){U^{QE}}^{\dagger}P^E\bigr) 
=\mbox{tr}_{RE}
\bigl(P^E U^{QE}(\rho^{RQ}\otimes\rho^E){U^{QE}}^{\dagger}P^E\bigr)\;.
\end{equation}

We denote the {\em normalized\/} states of the different systems $R$, 
$Q$, and $E$ {\em after\/} this evolution by primes.  Of special interest 
is the joint state of $RQ$ after the dynamics, which is given by 
\begin{equation}
\rho^{RQ'}=
\frac{\mbox{tr}_{E}
\bigl(P^E U^{QE}(\rho^{RQ}\otimes\rho^E){U^{QE}}^{\dagger}P^E\bigr)}
{\mbox{tr}
\bigl(P^E U^{QE}(\rho^{RQ}\otimes\rho^E){U^{QE}}^{\dagger}P^E\bigr)}
=\frac{
({\cal I}^R\otimes{\cal E})(\rho^{RQ})}
{\mbox{tr}
\bigl(({\cal I}^R\otimes{\cal E})(\rho^{RQ})\bigr)}
=\frac{
({\cal I}^R\otimes{\cal E})(\rho^{RQ})}
{\mbox{tr}
\bigl({\cal E}(\rho^{Q})\bigr)}
\;,
\label{eqtn: rhoRQprime}
\end{equation}
where ${\cal I}^R$ is the identity operation for the reference system.
Using the operator-sum decomposition (\ref{eqtn: gen op}) of ${\cal E}$, 
we can write $\rho^{RQ'}$ in terms of an ensemble of unnormalized pure 
states, that is, as a sum of terms each of which is proportional to a 
one-dimensional projection operator:
\begin{equation}
\rho^{RQ'}=
\sum_j
{(I^R\otimes A_j)\bigl|\Psi^{RQ}\bigl\rangle
\bigr\langle\Psi^{RQ}\bigr|(I^R\otimes A_j^\dagger)
\over
\mbox{tr}\bigl({\cal E}(\rho^Q)\bigr)}
\;.
\label{eqtn: rhoRQprimetwo}
\end{equation}

The state of the reference system after the dynamics is given by
\begin{equation}
\rho^{R'}=  
\frac{\mbox{tr}_{QE}
\bigl(P^E U^{QE}(\rho^{RQ}\otimes\rho^E){U^{QE}}^{\dagger}P^E\bigr)}
{\mbox{tr}
\bigl(P^E U^{QE}(\rho^{RQ}\otimes\rho^E){U^{QE}}^{\dagger}P^E\bigr)}
=\mbox{tr}_Q(\rho^{RQ'})
\;.
\label{eqtn: rhoRprime}
\end{equation}
We emphasize that $\rho^{R'}$ is generally not the same as $\rho^R$,
because of the presence of the environment projector $P^E$ in 
Eq.~(\ref{eqtn: rhoRprime}).  This is in contrast to the case of a
trace-preserving operation ${\cal E}$, for which the environment
projector is absent, that is, $P^E=I^E$ in Eq.~(\ref{eqtn: rhoRprime})
and, hence, for which $\rho^{R'}=\rho^R$.

\subsection{Entanglement fidelity and entropy exchange}
\label{sect: fid-ent}

{}Following Schumacher \cite{Schumacher96a,Schumacher96b}, we define
the {\em entanglement fidelity\/} to be the fidelity with which 
the joint state of $RQ$ is preserved by the dynamics:
\begin{equation}
{}F_e(\rho,{\cal E})\equiv
\big\langle\Psi^{RQ}\bigl|\rho^{RQ'}\bigr|\Psi^{RQ}\big\rangle
={\big\langle\Psi^{RQ}\bigl|
({\cal I}^R\otimes{\cal E})(\rho^{RQ})
\bigr|\Psi^{RQ}\big\rangle
\over
\mbox{tr}\bigl({\cal E}(\rho^Q)\bigr)}
\;.
\end{equation}
Using the form~(\ref{eqtn: rhoRQprimetwo}) of $\rho^{RQ'}$ and noting that
\begin{equation}
\bigl\langle\Psi^{RQ}\bigl|(I^R\otimes A_j)\bigr|\Psi^{RQ}\bigr\rangle=
\mbox{tr}\Bigl(
\bigr|\Psi^{RQ}\bigr\rangle\bigl\langle\Psi^{RQ}\bigl|
(I^R\otimes A_j)\Bigr)
=\mbox{tr}(\rho^Q A_j)
\;,
\end{equation}
we can put the entanglement fidelity in a form that, as implied
by our notation $F_e(\rho, {\cal E})$, manifestly depends only on
the initial state of $Q$ and the operation that is applied to $Q$,
\begin{equation}
{}F_e(\rho,{\cal E})=
\sum_j
{\bigl|\mbox{tr}(\rho A_j)\bigr|^2\over\mbox{tr}\bigl({\cal E}(\rho)\bigr)}\;.
\end{equation}

The {\em entropy exchange\/} is defined to be
\begin{equation}
S_e(\rho,{\cal E}) \equiv S(\rho^{RQ'}) = S(\rho^{E'})
\;, 
\end{equation}
where $S(\rho)=-\mbox{tr}(\rho\log\rho)$ denotes the von Neumann entropy 
of the density operator $\rho$ and where the latter equality follows from 
the fact that the overall state after the two-part dynamics is pure.
This generalizes the definition of entropy exchange in~\cite{Schumacher96a} 
to general quantum operations ${\cal E}$.

The entropy exchange obeys several inequalities that follow from the
subadditivity of von Neumann entropy \cite{Araki70a,Lieb75a,Wehrl78a} and 
the purity of the overall state of $RQE$ after the dynamics:
\begin{eqnarray}
S_e=
S(\rho^{RQ'})&\le&
S(\rho^{R'})+S(\rho^{Q'})\;,
\label{eqtn: subadditivityRQ}\\
S(\rho^{Q'})=
S(\rho^{RE'})&\le&
S(\rho^{R'})+S(\rho^{E'})=
S(\rho^{R'})+S_e\;,
\label{eqtn: subadditivityRE}\\
S(\rho^{R'})=
S(\rho^{QE'})&\le&
S(\rho^{Q'})+S(\rho^{E'})=
S(\rho^{Q'})+S_e\;.
\label{eqtn: subadditivityQE}
\end{eqnarray}
Each inequality here is an expression of subadditivity, with equality
holding if and only if the joint density operator on the left factors
into a product of the two density operators on the right (for example,
equality holds in Eq.~(\ref{eqtn: subadditivityRQ}) if and only if
$\rho^{RQ'}=\rho^{R'}\otimes\rho^{Q'}$).  The last two of the above 
inequalities can be combined into a single inequality, 
\begin{equation}
S_e(\rho,{\cal E})=
S(\rho^{RQ'})\ge
\bigl|S(\rho^{Q'})-S(\rho^{R'})\bigr|
\;,
\label{eqtn: ArakiLieb}
\end{equation}
sometimes known as the {\em Araki-Lieb inequality} \cite{Araki70a,Wehrl78a}.

If ${\cal E}$ is trace preserving, then as noted below 
Eq.~(\ref{eqtn: rhoRprime}), the state of the reference system does 
not change under the dynamics, that is, $\rho^{R'}=\rho^R$; moreover, 
since the initial state of $RQ$ is pure, we always have that 
$S(\rho^R)=S(\rho^Q)$.  Thus, for a trace-preserving operation, we
can use $S(\rho^{R'})=S(\rho^Q)$ to eliminate the reference system
from the above inequalities, leaving the inequalities obtained by
Schumacher \cite{Schumacher96a}.  For a general quantum operation,
it is not true that $S(\rho^{R'})=S(\rho^Q)$, and the inequalities
must be left in the form given above. 

The von Neumann entropy of $\rho^{RQ'}$ is the same as the entropy of
the matrix of inner products formed from the unnormalized pure states 
that contribute to the ensemble for $\rho^{RQ'}$ in 
Eq.~(\ref{eqtn: rhoRQprimetwo}).  Explicitly, in terms of a positive,
unit-trace matrix $W$, whose components are 
\begin{equation}
W_{jk}=
{\bigl\langle\Psi^{RQ}\bigl|(I^R\otimes A_k^\dagger)
(I^R\otimes A_j)\bigr|\Psi^{RQ}\bigr\rangle
\over
\mbox{tr}\bigl({\cal E}(\rho^Q)\bigr)}
\;,
\end{equation}
the entropy exchange is given by
\begin{equation}
S_e(\rho,{\cal E}) = S(W) = - \mbox{tr}(W \log W)\;. 
\end{equation}
The components of $W$ can be simplified to the form
\begin{equation}
W_{jk}=
{\mbox{tr}\Bigl(\bigl|\Psi^{RQ}\bigl\rangle\bigr\langle\Psi^{RQ}\bigr|
(I^R\otimes A_k^\dagger)(I^R\otimes A_j)\Bigr)
\over
\mbox{tr}\bigl({\cal E}(\rho^Q)\bigr)}
={\mbox{tr}(A_j\rho A_k^\dagger)
\over
\mbox{tr}\bigl({\cal E}(\rho)\bigr)}
\;,
\label{eqtn: Wmatrix}
\end{equation}
Notice that the diagonal elements of $W$,
\begin{equation}
q_j=W_{jj}= 
{\mbox{tr}(A_j\rho A_j^\dagger) \over \mbox{tr}\bigl({\cal E}(\rho)\bigr)}
\;,
\end{equation}
are the probabilities with which the pure states in 
Eq.~(\ref{eqtn: rhoRQprimetwo}) contribute to the ensemble.

Relative to a particular density operator $\rho$, there is a ``canonical 
decomposition'' of the quantum operation ${\cal E}$.  Suppose the matrix 
$W$ arises from $\rho$ and a particular operator-sum decomposition of 
${\cal E}$ in terms of decomposition operators $A_j$, as in 
Eq.~(\ref{eqtn: Wmatrix}), and suppose we diagonalize $W$ with a 
unitary matrix $u$,
\begin{equation}
\sum_{l,m} u_{jl} W_{lm} u_{km}^* = \lambda_j \delta_{jk}\;, 
\end{equation}
where the nonnegative real numbers $\lambda_j$ are the eigenvalues of $W$.
Now define new operators
\begin{equation}
\tilde A_j \equiv \sum_k u_{jk} A_k\;. 
\label{eqtn: canonical decomposition}
\end{equation}
These operators being a unitary remixing of the original decomposition
operators, they give another operator-sum decomposition for ${\cal E}$, 
with an associated matrix
\begin{equation} \label{eqtn: canonical conditions}
\tilde W_{jk} = 
\frac{\mbox{tr}(\tilde A_j \rho \tilde A_k^{\dagger})}
{\mbox{tr}\bigl({\cal E}(\rho)\bigr)}= 
\sum_{l,m} u_{jl} W_{lm} u_{km}^* = \lambda_j \delta_{jk}\;. 
\end{equation}

We say that a decomposition $\tilde A_j$ satisfying 
Eq.~(\ref{eqtn: canonical conditions}) is a {\em canonical decomposition\/}
of ${\cal E}$ with respect to $\rho$.  The canonical decomposition is 
unique up to degeneracies in the eigenvalues $\lambda_j$ and up to (trivial) 
phase changes in the canonical decomposition operators $\tilde A_j$.  
The entropy exchange can be written as
\begin{equation}
S_e=S(W)=S(\tilde W)=-\sum_j \lambda_j\log \lambda_j\equiv H(\vec\lambda\,)\;,
\end{equation}
where $H(\vec\lambda\,)$ is the Shannon information constructed from the
probability distribution $\lambda$.  It can be shown that
\begin{equation}
S_e\le H(\vec q\,)=-\sum_j q_j\log q_j\;, 
\label{eqtn: SEminShannon}
\end{equation}
equality holding only for a canonical decomposition.

Notice that for a pure quantum operation, the decomposition that 
contains only a single decomposition operator, as in 
Eq.~(\ref{eqtn: pureoperation}), is the canonical decomposition with
respect to any density operator $\rho$.  The canonical $W$ matrix is 
the one-dimensional unit matrix, and hence the entropy exchange $S_e$ 
is zero.

Consider the canonical decomposition of an operation ${\cal E}$ with 
respect to the unit density operator $I/D$, where $D$ is the dimension of 
the system Hilbert space.  Such a canonical decomposition, whose 
decomposition operators satisfy
\begin{equation}
\mbox{tr}(\tilde A_j\tilde A_k^\dagger)=
\lambda_j\mbox{tr}\bigl({\cal E}(I)\bigr)\delta_{jk}\;,
\end{equation}
is a minimal decomposition of ${\cal E}$.  All minimal decompositions can
be obtained from this one by unitary remixings that leave the number of
operators in the decomposition unchanged.

The entanglement fidelity and the entropy exchange obey the 
{\em quantum Fano inequality}, 
\begin{equation} \label{eqtn: qfano}
S_e(\rho,{\cal E}) \leq h\bigl(F_e(\rho,{\cal E})\bigr) + 
\bigl(1-F_e(\rho,{\cal E})\bigr)\log (D^2-1)\;,
\end{equation}
where $h(p) \equiv -p \log p - (1-p) \log (1-p)$ and $D$ is the
dimension of the system Hilbert space.  The quantum Fano inequality was 
first derived by Schumacher \cite{Schumacher96a} for trace-preserving 
operations, but Schumacher's proof goes through unchanged for general 
quantum operations.

\subsection{Data-processing inequality}
\label{sect: dataprocessing}

In the following we often consider a two-step operation that consists
of two successive operations.  The situation of interest here is that
of reversing an operation, as discussed in Sec.~\ref{sect: reversal}: 
the first of the two operations is the operation to be reversed, and 
the second is a reversal operation, which is necessarily deterministic.  
Schumacher and Nielsen \cite{Schumacher96b} derived an important 
inequality, called the data-processing inequality, for the case in 
which both operations in a two-step operation are trace preserving.  
Here we show that the data-processing inequality remains valid for 
any two-step operation in which the first operation ${\cal E}$ is 
arbitrary, but the second operation ${\cal D}$ is deterministic. 

In this situation we use $E$ to denote the environment used in the 
first step and $A$ to denote the ancilla or environment used in the 
second step.  We use double primes to distinguish the normalized 
states of the various systems after the second step.  We can draw 
several conclusions about the von Neumann entropies of various states 
in this scenario.  In particular, since the overall state after both 
steps is pure, we have that $S(\rho^{REA''})=S(\rho^{Q''})$ and that
$S(\rho^{EA''})=S(\rho^{RQ''})=S_e(\rho,{\cal D}\circ{\cal E})$ is the
entropy exchange for the two-step operation.  Moreover, since the 
state of $RQE$ is pure after the first step and since $R$ and $E$ do 
not participate in the second step, we have that 
$S(\rho^{RE''})=S(\rho^{RE'})=S(\rho^{Q'})$ and that 
$S(\rho^{E''})=S(\rho^{E'})=S(\rho^{RQ'})=S_e(\rho,{\cal E})$ is the
entropy exchange in the first step. 

The strong subadditivity property of von Neumann entropy 
\cite{Lieb75a,Wehrl78a,Lieb73a} constrains the entropies after the 
two-step dynamics:
\begin{equation}
S(\rho^{REA''})+S(\rho^{E''})\le
S(\rho^{RE''})+S(\rho^{EA''})\;.
\end{equation}
Substituting the entropy relations just derived and re-arranging
yields the {\em data-processing inequality},
\begin{equation}
S(\rho^{Q'})-S_e(\rho,{\cal E})\ge
S(\rho^{Q''})-S_e(\rho,{\cal D}\circ{\cal E})
\;.
\end{equation}
The left-hand side of the data-processing inequality is constrained
by Eq.~(\ref{eqtn: subadditivityRE}), leading to the double 
inequality
\begin{equation}
S(\rho^{R'})\ge
S(\rho^{Q'})-S_e(\rho,{\cal E})\ge
S(\rho^{Q''})-S_e(\rho,{\cal D}\circ{\cal E})
\label{eqtn: fulldataprocessing}
\;.
\end{equation}
If ${\cal E}$ is trace preserving, then $S(\rho^{R'})=S(\rho^Q)$ and
this double inequality reduces to the form found by Schumacher and
Nielsen \cite{Schumacher96b}.

\section{Information-theoretic characterization of reversible quantum 
operations}

\subsection{General information-theoretic characterization}
\label{sect: information characterization}

In this section we demonstrate that a general quantum operation $\cal E$ 
is reversible on a subspace $M$ of the total state space $L$ if and only 
if the following two conditions are satisfied:
\begin{equation}
\mbox{\it Condition~1:}\quad 
\mbox{tr}\bigl({\cal E}(\rho)\bigr) = \mu^2\quad
\mbox{for all $\rho$ whose support is confined to $M$,} 
\label{eqtn: condition1}
\end{equation}
where $\mu$ is a real constant satisfying $0 < \mu \leq 1$;
\begin{eqnarray}
&\mbox{}&\mbox{\it Condition~2:}\nonumber\\
&\mbox{}&\qquad
S(\rho)=
S\!\left({{\cal E}(\rho)\over\mbox{tr}\bigl({\cal E}(\rho)\bigr)}\right) 
- S_e(\rho,{\cal E})\quad
\mbox{for any {\it one\/} $\rho$ whose support is 
the {\it entirety\/} of $M$,}\nonumber\\
\label{eqtn: condition2}
\end{eqnarray}
(and then for all $\rho$ whose support is confined to $M$).

Condition~1 is equivalent to 
\begin{equation}
P_M E P_M = \mu^2 P_M\;, 
\label{eqtn: condition1prime}
\end{equation}
where 
\begin{equation}
E\equiv\sum_j A_j^{\dagger} A_j
\label{eqtn: POVME}
\end{equation}
is the POVM element corresponding to ${\cal E}$ and $P_M$ is the projector 
onto $M$.  Condition~1 has the appealing intuitive interpretation that 
if we view ${\cal E}$ as a dynamics for the system, conditional on some 
measurement result or post-interaction environment state, knowledge of 
that result or state gives no information about the initial system state 
$\rho$.  Condition~2, though less intuitive, states essentially that for 
initial states whose support lies in $M$, no quantum information is lost 
in the dynamics described by ${\cal E}$.

We begin by proving necessity.  Suppose that ${\cal E}$ is reversible 
on $M$.  Then it was shown in~\cite{Nielsen96b} that Condition~1 follows.  
The reason is that reversibility implies that 
${\cal R}\circ{\cal E}(\rho) = \mbox{tr}\bigl({\cal E}(\rho)\bigr) \rho$
for all $\rho$ whose support is confined to $M$; ${\cal R}\circ{\cal E}$ 
being linear, the only way this equation can be satisfied is if 
$\mbox{tr}({\cal E}(\rho))$ has a constant value $\mu^2>0$.  Let 
${\cal E}_M$ be the restriction of ${\cal E}$ to $M$, that is,
\begin{equation}
{\cal E}_M(\rho) \equiv \sum_j A_j P_M \rho P_M A_j^{\dagger}\;.
\label{eqtn: EsubM}
\end{equation}
Notice that ${\cal E}_M(\rho) = {\cal E}(\rho)$ if $\rho$ has support 
lying wholly in $M$.  Let $\overline M$ be the subspace that is the 
orthocomplement of $M$ and $P_{\,\overline M\,}$ be the projector onto
$\overline M$.  Now introduce a new quantum operation ${\cal F}$, whose 
action on {\it any\/} $\rho$ is given by
\begin{equation}
{\cal F}(\rho) \equiv \frac{{\cal E}_M(\rho)}{\mu^2} + 
P_{\,\overline M\,}\rho P_{\,\overline M\,}\;.
\end{equation}
The reason for introducing ${\cal F}$ is that it is a deterministic
operation with the property that 
\begin{equation} \label{eqtn: fande}
{\cal F}(\rho)=
\frac{{\cal E}(\rho)}{\mu^2}=
\frac{{\cal E}(\rho)}{\mbox{tr}\bigl({\cal E}(\rho)\bigr)}
\end{equation}
for states $\rho$ whose support lies wholly in $M$.  Thus ${\cal E}$ is 
reversible on $M$ if and only if ${\cal F}$ is reversible on $M$. Since 
${\cal F}$ is deterministic, however, the necessary and sufficient 
condition for its reversibility is the condition already obtained by 
Schumacher and Nielsen \cite{Schumacher96b}: ${\cal F}$ is reversible on 
$M$ if and only if
\begin{equation} \label{eqtn: schuniels}
S(\rho) = S\bigl({\cal F}(\rho)\bigr) - S_e(\rho,{\cal F})
\end{equation}
for any one $\rho$ whose support is the entirety of $M$ (and then for
all $\rho$ whose support lies in $M$).  We complete the proof of necessity 
by noting that for states $\rho$ whose support is confined to $M$, the 
$W$ matrices of ${\cal E}$ and ${\cal F}$ are the same, which implies that 
\begin{equation} \label{eqtn: seforfande}
S_e(\rho,{\cal E}) = S_e(\rho,{\cal F})\;. 
\end{equation}
Substituting Eqs.~(\ref{eqtn: seforfande}) and (\ref{eqtn: fande}) into 
Eq.~(\ref{eqtn: schuniels}) yields the second 
condition~(\ref{eqtn: condition2}).

The sufficiency of Conditions~1 and 2 is proved in an obviously similar 
way, but one point should be stressed.  For ${\cal F}$ to be reversible 
on $M$, it is sufficient that Eq.~(\ref{eqtn: schuniels}) hold for any 
{\it one\/} $\rho$ whose support is the {\it entirety\/} of $M$.  Thus 
for ${\cal E}$ to be reversible, it is sufficient that the second 
condition~(\ref{eqtn: condition2}) hold for any one such $\rho$.  

Before going on, one further point deserves mention.  If the initial
state of $Q$ is the unit density operator in the subspace $M$---that is,
$\rho=P_M/d$, where $d\le D$ is the dimension of $M$---then we can 
dispense with Condition~1.  What we are claiming is the following 
equivalence: ${\cal E}$ is reversible on $M$ if and only if 
\begin{equation}
\log d=S(P_M/d) = 
S\!\left({{\cal E}(P_M/d)\over\mbox{tr}\bigl({\cal E}(P_M/d)\bigr)}\right) 
- S_e(P_M/d,{\cal E})\;.
\label{eqtn: condition2prime}
\end{equation}
The necessity of Eq.~(\ref{eqtn: condition2prime}) has already been 
shown.  We now demonstrate sufficiency by showing that 
Eq.~(\ref{eqtn: condition2prime}) implies the first 
condition~(\ref{eqtn: condition1}).  

{}For this purpose, notice that when $\rho^Q=P_M/d$, the initial pure state
of $RQ$ is an entangled state of the form
\begin{equation}
\bigl|\Psi^{RQ}\bigr\rangle=
{1\over\sqrt d}\sum_{m=1}^d |\chi_m^R\rangle\otimes|\phi_m^Q\rangle
\;.
\end{equation}
Here the kets $|\phi_m^Q\rangle$ are an orthonormal basis for the 
$d$-dimensional subspace $M$, and the kets $|\chi_m^R\rangle$ are a set 
of $d$ orthonormal vectors for $R$.  Substituting this entangled state 
into Eq.~(\ref{eqtn: rhoRQprimetwo}) yields a new expression for the 
joint state of $RQ$ after the dynamics:
\begin{equation}
\rho^{RQ'}=
\sum_{m,n}|\chi_m^R\rangle\langle\chi_n^R|\otimes
{1\over d}\sum_j
{A_j|\phi_m^Q\rangle\langle\phi_n^Q|A_j^\dagger
\over
\mbox{tr}\bigl({\cal E}(P_M/d)\bigr)}
\;.
\end{equation}
The state of the reference system after the dynamics now assumes the form
\begin{equation}
\rho^{R'}=
\mbox{tr}_Q(\rho^{RQ'})=
\sum_{m,n}
{1\over d}
{\langle\phi_n^Q|E|\phi_m^Q\rangle
\over
\mbox{tr}\bigl({\cal E}(P_M/d)\bigr)}\,
|\chi_m^R\rangle\langle\chi_n^R|
\;,
\label{eqtn: rhoRprimetwo}
\end{equation}
where $E$ is the POVM element associated with ${\cal E}$ [cf.\ 
Eq.~(\ref{eqtn: POVME})].  Since the support of $\rho^{R'}$ lies 
within the $d$-dimensional subspace spanned by the vectors 
$|\chi_m^R\rangle$, we know that $S(\rho^{R'})\le\log d$.  
Condition~(\ref{eqtn: condition2prime}), when combined with the left 
inequality in Eq.~(\ref{eqtn: fulldataprocessing}), implies that 
$S(\rho^{R'})=\log d$.  This means that the matrix elements of 
$\rho^{R'}$ in Eq.~(\ref{eqtn: rhoRprimetwo}) must be those of the
unit density operator $P_M/d$ on $M$.  Hence we conclude that
\begin{equation}
P_M EP_M=
\mbox{tr}\bigl({\cal E}(P_M/d)\bigr)P_M
\;,
\end{equation}
which as already noted in Eq.~(\ref{eqtn: condition1prime}), is equivalent 
to Condition~1, with $\mu^2=\mbox{tr}\bigl({\cal E}(P_M/d)\bigr)$.

\subsection{Applications of information-theoretic conditions for 
reversibility}
\label{sect: applications}

A number of useful results follow from the information-theoretic 
characterization of reversible quantum operations found in the preceding 
subsection.  Among these is a general characterization of teleportation 
schemes, which has been discussed in~\cite{Nielsen96b} and is the 
subject of continuing work.  This subsection describes several simpler, 
but still useful applications of the information-theoretic characterization.

We first show that an operation is reversible by a unitary operation
if and only if it acts like a multiple of a unitary operation when 
restricted to the reversal subspace.

{\em Theorem}.  A quantum operation ${\cal E}$, with decomposition
operators $A_j$, is reversible by a unitary operator $U$ on a subspace 
$M$ if and only if there exist complex constants $c_j$ such that
\begin{equation}
A_j P_M = c_j U^{\dagger} P_M\;. 
\label{eqtn: condunitary}
\end{equation}
The constants satisfy 
\begin{equation}
\sum_j|c_j|^2=\mu^2\;,
\end{equation}
where $\mu^2$ is the constant value of $\mbox{tr}({\cal E}(\rho))$ on $M$.

{\em Proof}.  The sufficiency of the condition~(\ref{eqtn: condunitary})
is obvious.  The proof of necessity is to notice that for all $\rho$, 
not just those whose support is confined to $M$, we have
\begin{equation}
U{{\cal E}_M(\rho)\over\mu^2}U^\dagger=P_M\rho P_M
\;,
\end{equation}
where ${\cal E}_M$ is the restriction of $\rho$ to $M$.  Rewritten as
\begin{equation}
{\cal E}_M(\rho)=
\sum_j A_j P_M \rho P_M A_j^{\dagger} = 
\mu^2 U^{\dagger} P_M \rho P_M U\;,
\end{equation}
this shows that ${\cal E}_M$ is a pure operation whose canonical 
decomposition contains the single operator $\mu U^\dagger P_M$.  By 
the result~(\ref{eqtn: decompositions}) that relates operator-sum 
decompositions, the conclusion follows.  This completes the proof.  

A second theorem shows that an operation that is reversible for all 
initial states acts like a multiple of a unitary operation.  

{\em Theorem}.  A quantum operation ${\cal E}$ that is reversible on 
the entire state space of the system is a positive multiple of a 
unitary operation; that is,
\begin{equation}
{\cal E}(\rho) = \mu^2 U \rho U^{\dagger}
\end{equation}
for some constant $\mu$ satisfying $0<\mu\le1$.

{\em Proof}.  A simple proof can be obtained by examining the reversal 
operation constructed in~\cite{Schumacher96b} and verifying that it 
is unitary. The result follows from this and the fact that 
$\mu^2\equiv\mbox{tr}({\cal E}(\rho))$ is a constant.

We present here, however, a purely information-theoretic proof that does 
not require the explicit construction of a reversal operation.  A quantum 
operation that is reversible on the entire $D$-dimensional state space of 
$Q$ satisfies
\begin{equation}
\log D=
S(I/D)=
S\!\left(\frac{{\cal E}(I/D)}{\mu^2}\right) - S_e(I/D,{\cal E})\;.
\end{equation}
Since the entropy is maximized by $I/D$ and the entropy exchange 
is nonnegative, we see immediately that 
$S\bigl({\cal E}(I/D)/\mu^2\bigr)=\log D$ and $S_e(I/D,{\cal E})=0$.  
The first of these conclusions means that
\begin{equation}
{\cal E}(I/D)=\mu^2(I/D)\;.
\label{eqtn: EID}
\end{equation}
The second means that the $W$ matrix for initial state $I/D$ is of 
rank one; thus there exists a unitary matrix $u$ such that
\begin{equation}
\sum_{l,m} u_{jl} W_{lm} u_{km}^* = \delta_{j1}\delta_{k1}
\;. 
\end{equation}
Defining a canonical decomposition of ${\cal E}$ as in 
Eq.~(\ref{eqtn: canonical decomposition}), the canonical $W$ 
matrix becomes
\begin{equation}
\tilde W_{jk} = 
\frac{\mbox{tr}(\tilde A_j\tilde A_k^{\dagger})}{D\mu^2}=
\delta_{j1}\delta_{k1}
\;.
\end{equation}
It follows that only the first operator in the canonical decomposition,
$\tilde A_1$, is nonzero.  Combining this result with Eq.~(\ref{eqtn: EID}) 
yields
\begin{equation}
{\cal E}(I/D)=\tilde A_1\tilde A_1^\dagger/D=\mu^2(I/D)\;,
\end{equation}
which implies that $\tilde A_1=\mu\,U$.  This completes the proof.

This second result has a useful application to teleportation.  Recall 
the basic set-up for teleportation \cite{Bennett93a}.  Alice possesses
an unknown quantum state $\rho$, which she wishes to 
teleport to Bob.  Alice also sends Bob some classical information, 
which we represent by $i$.  It was shown in~\cite{Nielsen96b} that 
the state of Bob's system, conditioned on the information $i$, is 
related to the state of Alice's system by a quantum operation 
${\cal E}_i$,
\begin{equation}
\rho \rightarrow \frac{{\cal E}_i(\rho)}{\mbox{tr}({\cal E}_i(\rho))}.
\end{equation}
If Bob wishes to achieve teleportation then he must be able to reverse
the operation ${\cal E}_i$.  What the above result shows is that Bob can 
use a unitary operation to do the reversal, since his reversal must work 
over the entire space of initial states $\rho$.  No generality is 
introduced by allowing Bob to use nonunitary reversal operations, that 
is, by allowing Bob to employ an ancilla to assist in teleportation.  
Considering only unitary reversals, as was done in~\cite{Nielsen96b}, is 
thus sufficient for the study of teleportation.

We can also compare the results obtained in this paper to earlier 
characterizations of reversible pure and deterministic quantum 
operations, obtained in~\cite{Schumacher96b} and~\cite{Nielsen96b} 
(where pure operations are called ideal operations), and show that 
these earlier results are special cases of the general characterization 
embodied in Conditions~1 and 2.

{}For a deterministic quantum operation ${\cal E}$, it is certainly true 
that $\mbox{tr}\bigl({\cal E}(\rho)\bigr) = 1$ is a constant, so 
Condition~1 is automatic.  Thus reversibility for a deterministic 
operation is equivalent to Condition~2 alone, that is, to 
$S(\rho)=S\bigl({\cal E}(\rho)\bigr)- S_e(\rho,{\cal E})$, which is 
the reversibility condition obtained in~\cite{Schumacher96b}. 

{}For a pure quantum operation ${\cal E}$, Condition~1 is equivalent to 
$P_MA^\dagger AP_M=\mu^2P_M$.  This implies, using the polar-decomposition
property of operators, that 
$AP_M=U^\dagger\sqrt{P_M A^\dagger AP_M}=\mu U^\dagger P_M$ for some unitary 
operator $U$.  Hence ${\cal E}$ can be reversed by $U$.  Thus reversibility 
for a pure operation is equivalent to Condition~1 alone, that is, to
$\mbox{tr}\bigl({\cal E}(\rho)\bigr)=\mu^2$, which is the reversibility
condition obtained in~\cite{Nielsen96b}.  For a pure operation,
Condition~2 can be dispensed with because it follows from Condition~1: 
for any pure operation the entropy exchange is zero, and Condition~1 
implies that ${\cal E}$ acts like a multiple of a unitary on $M$, which 
means that $S(\rho)=S\bigl({\cal E}(\rho)/\mu^2\bigr)$.

\section{Algebraic characterization of reversible operations}
\label{sect: algebraic reversal}

Up to this point we have taken an information-theoretic approach to
the reversal of quantum operations.  In this section we switch to an 
algebraic point of view.  The algebraic results obtained in this section 
are particularly powerful when used in combination with the 
information-theoretic viewpoint, as we illustrate in the next section 
on the thermodynamic cost of error correction.  We begin with the 
theorem that establishes algebraic conditions for reversibility.

\subsection{Reversibility theorem}
\label{sect: algebraic theorem}

{\em Theorem}.  A quantum operation ${\cal E}$, with decomposition 
operators $A_j$, is reversible on $M$ if and only if there exists a 
positive matrix $m$ such that
\begin{equation} \label{eqtn: algebraic condition}
P_M A_k^{\dagger} A_j P_M = m_{jk} P_M\;. 
\end{equation}
The trace of $m$,
\begin{equation}
\sum_j m_{jj}=\mu^2\;,
\end{equation}
is the constant value of $\mbox{tr}\bigl({\cal E}(\rho)\bigr)$ on 
$M$.  (Under a unitary remixing of the decomposition operators, 
the matrix $m$ undergoes a unitary transformation, which leaves 
the trace invariant.)

This result was proved by Knill and Laflamme \cite{Knill97a} and by 
Bennett {\em et al.}\ \cite{Bennett96a}.  We give a different proof, 
particularly of the sufficiency of 
condition~(\ref{eqtn: algebraic condition}).  The construction used in 
our proof is crucial to our subsequent analysis of the thermodynamics 
of error correction.

{\em Proof}.  We deal first with the necessity of 
condition~(\ref{eqtn: algebraic condition}) and notice that for all 
density operators, not just those whose support is confined to $M$,
we have 
\begin{equation}
{\cal R}\circ{\cal E}_M(\rho)=
\sum_{l,j}R_lA_jP_M\rho P_MA_j^\dagger R_l^\dagger=
\mu^2 P_M\rho P_M
\;,
\label{eqtn: reversalcondition2}
\end{equation}
where the operators $R_k$ make up an operator-sum decomposition for the 
reversal operation ${\cal R}$.  Equation~(\ref{eqtn: reversalcondition2}) 
means that ${\cal R}\circ{\cal E}_M$ is a pure operation, whose canonical 
decomposition consists of the single operator $\mu P_M$.  By the 
result~(\ref{eqtn: decompositions}) that relates operator-sum 
decompositions, we can conclude that there exist constants $c_{jl}$ 
such that
\begin{equation}
R_lA_jP_M=c_{jl}P_M\;.
\label{eqtn: RAP}
\end{equation}
The constants satisfy
\begin{equation}
\sum_{j,l}|c_{jl}|^2=\mu^2\;.
\end{equation}
Using the trace-preserving property of the reversal operation ${\cal R}$, 
we can write
\begin{equation}
P_M A_k^{\dagger} A_j P_M= 
\sum_l P_M A_k^{\dagger}R_l^\dagger R_l A_j P_M 
=\left(\sum_l c_{jl}c_{kl}^*\right)\! P_M
=m_{jk}P_M\;, 
\end{equation}
where the matrix $m=cc^\dagger$ is manifestly positive.

We now demonstrate that condition~(\ref{eqtn: algebraic condition}) is
sufficient for reversibility.  Let $u$ be a unitary matrix that 
diagonalizes $m$, that is, 
\begin{equation}
\sum_{l,n} u_{jl} m_{ln} u_{kn}^* = d_j \delta_{jk}\;, 
\label{eqtn: diagonal m}
\end{equation}
where the nonnegative real numbers $d_j$ are the nonnegative eigenvalues of 
$m$.  Relative to a new decomposition of ${\cal E}$, defined by
\begin{equation}
\tilde A_j \equiv \sum_j u_{jk} A_k\;, 
\label{eqtn: canonical for reversible}
\end{equation}
condition~(\ref{eqtn: algebraic condition}) becomes
\begin{equation} \label{eqtn: canonical error correction}
P_M \tilde A_k^{\dagger} \tilde A_j P_M = d_j \delta_{jk} P_M\;.
\end{equation}
The diagonal ($j=k$) elements of Eq.~(\ref{eqtn: canonical error correction})
imply, by the polar-decomposition property, that there exist unitary 
operators $U_j$ such that
\begin{equation}
\label{eqtn: restrictedpolar}
\tilde A_j P_M = U_j \sqrt{P_M \tilde A_j^{\dagger} \tilde A_j P_M}
 	 = \sqrt{d_j} U_j P_M\;.
\end{equation}
Notice that if $d_j=0$, the corresponding decomposition operator
$\tilde A_j$ is irrelevant to the operation of ${\cal E}$ within $M$,
although such an $\tilde A_j$ is generally important to the action of 
${\cal E}$ on density operators whose support is not confined to $M$.  
When there are such decomposition operators, i.e., when $d_j=0$ for
some $j$, the subspace $M$ is called a {\em degenerate code\/}; we 
discuss the meaning and significance of degenerate codes in 
Sec.~\ref{sect: algebraic discussion}.  For $d_j\ne0$ we let $M_j$ 
be the subspace that $M$ is mapped to by $U_j$, and we let 
$P_j \equiv U_j P_M U_j^{\dagger}$ be the projector onto $M_j$.  The 
off-diagonal elements of Eq.~(\ref{eqtn: canonical error correction}) 
imply that these subspaces are orthogonal, that is, 
\begin{equation}
P_k P_j = \delta_{jk} P_j\;.
\end{equation}

The action of ${\cal E}$ on any density operator whose support is 
confined to $M$ takes the following form in terms of the new decomposition:
\begin{equation}
{\cal E}(\rho)=
\sum_j\tilde A_jP_M\rho P_M\tilde A_j^\dagger
=\sum_j d_jU_jP_M\rho P_MU_j^\dagger
=\sum_j d_jP_jU_j\rho U_j^\dagger P_j\;.
\label{eqtn: E on M}
\end{equation}
It is easy now to construct an operation that reverses ${\cal E}$. 
Let $N$ be the subspace that is the direct sum of the orthogonal
subspaces $M_j$, and let $\overline N$ be the orthocomplement of $N$.
The projector onto $\overline N$ is given by 
\begin{equation}
P_{\,\overline N\,} = I-P_N=I-\sum_{\{j\mid d_j\ne0\}} P_j\;.
\end{equation}
Now we define the action of a putative reversal operation by 
\begin{equation} 
\label{eqtn: reversalop}
{\cal R}(\rho) \equiv 
\sum_{\{j\mid d_j\ne0\}}U_j^{\dagger} P_j \rho P_j U_j 
+ P_{\,\overline N\,}\rho P_{\,\overline N\,}\;.
\end{equation}
This reversal operation is trace preserving, as required, since
\begin{equation}
\sum_{\{j\mid d_j\ne0\}} P_j U_j U_j^{\dagger} P_j + 
P_{\,\overline N\,} P_{\,\overline N\,}=  
\sum_{\{j\mid d_j\ne0\}} P_j + P_{\,\overline N\,} = I\;, 
\end{equation}
and simple algebra shows that for all $\rho$ whose support is confined
to $M$, 
\begin{equation}
{\cal R}\circ{\cal E}(\rho) = \mu^2 \rho\;, 
\end{equation}
where 
\begin{equation}
\mu^2=\sum_j d_j=\mbox{tr}\bigl({\cal E}(\rho)\bigr)\;.
\end{equation}
Thus ${\cal R}$ is indeed a reversal operation for ${\cal E}$ on the 
subspace $M$.  This completes the proof.

\subsection{Discussion of algebraic conditions for reversibility}
\label{sect: algebraic discussion}

The proof has a compelling physical interpretation.  It shows that an 
operation that is reversible on $M$ has an operator-sum decomposition 
in which the decomposition operators $\tilde A_j$ map $M$ 
{\em unitarily\/} to {\em orthogonal\/} subspaces $M_j$.  The operation 
on the entire space is generally not representable as an ensemble 
of unitary operations, but as far as its action on the reversible 
subspace $M$ is concerned, the operation {\em can\/} be represented 
by unitary operators $U_j$, which are applied randomly with probabilities 
$\lambda_j=d_j/\mu^2$ and which, moreover, take $M$ to orthogonal 
subspaces $M_j$.   Reversal can be effected by first measuring in 
which of the orthogonal subspaces the state lies after the operation, 
thus determining which unitary operator $U_j$ occurred, and then applying 
the corresponding inverse unitary operator $U_j^{\dagger}$ to restore 
the initial state.  We stress that one can {\em always\/} effect 
reversal in this way, by using a measurement---indeed, a pure, 
projection-valued measurement---followed by a unitary conditioned
on the result of the measurement.  It is equally important to emphasize,
however, that the deterministic reversal operation can also be constructed 
without measurements, by using an ancilla as described in 
Sec.~\ref{sect: qop}.  The two methods of reversal lead, of course, to 
the same reversal operation.

The operator-sum decomposition $\tilde A_j$ used in the above proof is 
obviously quite special.  It is a canonical decomposition for ${\cal E}$ 
relative to the initial state $P_M/d$, where $d$ is the dimension of $M$, 
as can be seen directly by taking the trace of 
Eq.~(\ref{eqtn: canonical error correction}).  More interesting is that 
the operators $\tilde A_j$ are a canonical decomposition for ${\cal E}$ 
relative to {\em any\/} initial density operator $\rho$ whose support 
lies in $M$; that is, the $W$ matrix is diagonal, 
\begin{equation}
\tilde W_{jk}=
\frac{\mbox{tr}(\tilde A_j\rho\tilde A_k^{\dagger})}{\mu^2}
=\frac{\mbox{tr}(\tilde A_jP_M\rho P_M\tilde A_k^{\dagger})}{\mu^2}=
{d_j\over\mu^2}\delta_{jk}=\lambda_j\delta_{jk}\;,
\end{equation}
with eigenvalues $\lambda_j$.  It follows that 
\begin{equation}
S_e(\rho,{\cal E})=S(\tilde W)=H(\vec\lambda)\;.
\end{equation}
Moreover, for any $\rho$ whose support lies in $M$, 
Eq.~(\ref{eqtn: E on M}) shows that the density operator after application 
of ${\cal E}$ is given by
\begin{equation}
\rho'=
{\cal E}(\rho)/\mu^2=
\sum_j\lambda_j\rho_j\;,
\label{eqtn: rhoprime}
\end{equation}
where
\begin{equation}
\rho_j\equiv U_j\rho U_j^\dagger
\end{equation}
is the unitary image of $\rho$ in $M_j$.  It follows that 
$S(\rho_j)=S(\rho)$ and since the density operators $\rho_j$ are 
orthogonal, that 
\begin{equation}
S\bigl({\cal E}(\rho)/\mu^2\bigr)=
H(\vec\lambda)+S(\rho)=
S(\rho)+S_e(\rho,{\cal E})\;.
\label{eqtn: condition2again}
\end{equation}
This is an explicit demonstration of condition~(\ref{eqtn: condition2}).

The existence of the canonical decomposition $\tilde A_j$ of 
Eq.~(\ref{eqtn: canonical for reversible}) clarifies the notion of a
degenerate code.  A common way of defining degeneracy is to say that a 
code is degenerate if any of the off-diagonal elements of the matrix 
$m$ in Eq.~(\ref{eqtn: algebraic condition}) are nonzero.  This 
definition is flawed, however, because it is not invariant under 
changes in the operator-sum decomposition of ${\cal E}$.  The 
off-diagonal elements of $m$ can always be made to vanish by 
transforming to the canonical decomposition $\tilde A_j$.  

Loosely speaking, what degeneracy is supposed to capture is the idea 
that some of the ``errors'' produced by ${\cal E}$ are irrelevant within 
the code subspace $M$.  This idea must be translated into a mathematical 
form that is independent of the operator-sum decomposition of ${\cal E}$.  
One way of doing so was introduced by Gottesman \cite{Gottesman96a}: 
suppose that the operators $A_j$ constitute a minimal (and thus linearly 
independent) decomposition of ${\cal E}$; if the restricted operators 
$A_jP_M$, which form a decomposition of the restricted operation 
${\cal E}_M$, are linearly dependent, Gottesman calls the code degenerate.  
The reason for this definition is that if the operators $A_jP_M$ are 
linearly dependent, then transformation to a minimal decomposition of 
${\cal E}_M$ reduces the number of decomposition operators, i.e., reduces 
the number of ``errors.''  The canonical decomposition provides just such 
a minimal decomposition of ${\cal E}_M$, that is, the decomposition 
consisting of the restricted operators $\tilde A_jP_M$.  The reduction 
in the number of errors shows up in that the operators $\tilde A_j$ that 
have $d_j=0$ are irrelevant to the operation of ${\cal E}$ within $M$.  
Thus we arrive at the equivalent definition of degeneracy introduced 
in the above proof: a code is degenerate if one or more of the eigenvalues 
$d_j$ vanishes.  

This discussion leads to a manifestly invariant way of defining degeneracy
in terms of operator subspaces: a code is degenerate if the operator 
subspace spanned by the decomposition operators of ${\cal E}$ has higher 
dimension than the operator subspace spanned by the decomposition operators 
of ${\cal E}_M$.  Moreover, it is now clear why degeneracy is considered 
a possible means of beating the ``quantum Hamming bound.''  That bound 
is derived from counting the number of possible linearly independent 
errors, not restricted to the code subspace, and assuming that error 
correction requires for each error an orthogonal subspace the same size 
as the reversible subspace $M$.


We turn now to properties of the reversal operation.  In 
Eq.~(\ref{eqtn: reversalop}) we introduced a particular reversal
operation ${\cal R}$, which is defined in terms of an operator-sum 
decomposition that consists of the operators 
\begin{equation}
\tilde R_j=U_j^\dagger P_j=P_M U_j^\dagger\quad
\mbox{for $j$ such that $d_j\ne0$,} 
\label{eqtn: reversal decompositionj}
\end{equation}
and the operator 
\begin{equation}
\tilde R_{\,\overline N\,}=P_{\,\overline N\,}\;.  
\label{eqtn: reversal decompositionN}
\end{equation}
The important part of ${\cal R}$ is its restriction to the subspace 
$N$; the action of the restricted operation ${\cal R}_N$ is defined by
\begin{equation}
{\cal R}_N(\rho)\equiv
\sum_j\tilde R_jP_N\rho P_N\tilde R_j^\dagger+
\tilde R_{\,\overline N\,}P_N\rho P_N\tilde R_{\,\overline N\,}
=\sum_j\tilde R_j\rho\tilde R_j^\dagger
=\sum_{\{j\mid d_j\ne0\}}U_j^{\dagger} P_j \rho P_j U_j 
\;.
\end{equation}

The first question we address is the extent to which the reversal
operation is unique.  For that purpose, consider another operation 
${\cal T}$, with decomposition operators $T_l$, which reverses ${\cal E}$
on $M$.  The action of ${\cal T}$ on an arbitrary density operator can 
be written as
\begin{equation}
{\cal T}(\rho)=
\sum_lT_lP_N\rho P_NT_l^\dagger+
\sum_l T_lP_{\,\overline N\,}\rho P_{\,\overline N\,}T_l^\dagger
+\sum_l T_l(P_N\rho P_{\,\overline N\,}+
P_{\,\overline N\,}\rho P_N)T_l^\dagger
\;.
\end{equation}
The first and second terms in this expression are the restrictions of 
${\cal T}$ to the orthogonal subspaces $N$ and $\overline N$, respectively,
and the third term is an additional contribution that can arise when 
$\rho$ is not block-diagonal with respect to $N$ and $\overline N$.  
The second and third terms are unaffected by the requirement that 
${\cal T}$ be a reversal operation; the only restrictions on the second 
and third terms come from the requirement that ${\cal T}$ be trace 
preserving.  The first term defines the action of the restriction of 
${\cal T}$ to the subspace $N$, that is,
\begin{equation}
{\cal T}_N(\rho)\equiv
\sum_lT_lP_N\rho P_NT_l^\dagger\;,
\end{equation}
The restricted operation ${\cal T}_N$ is the important part of ${\cal T}$
for reversal.  What we show now is that ${\cal T}_N$ is the same operation
as ${\cal R}_N$.

We proceed by noting that the decomposition operators $T_l$ must satisfy 
Eq.~(\ref{eqtn: RAP}) for any decomposition of ${\cal E}$ and, in particular,
must satisfy it when the decomposition operators for ${\cal E}$ are 
chosen to be the canonical decomposition $\tilde A_j$; that is, we
must have
\begin{equation}
\sqrt{d_j}T_lU_jP_M=T_l\tilde A_jP_M=\tilde c_{jl}P_M\;,
\end{equation}
where the constants $\tilde c_{jl}$ determine the diagonalized $m$ matrix
of Eq.~(\ref{eqtn: diagonal m}), 
\begin{equation}
\sum_l \tilde c_{jl}\tilde c_{kl}^*=d_j\delta_{jk}\;.
\end{equation}
We now discard the values of the index $j$ for which $d_j=0$; this 
eliminates rows of zeroes from the matrix $\tilde c$.  For the 
remaining values of $j$, we have that
\begin{equation}
T_lP_jU_j=T_lU_jP_M=
{\tilde c_{jl}\over\sqrt{d_j}}P_M
\equiv v_{lj}P_M\;,
\label{eqtn: TPU}
\end{equation}
The columns of the matrix $v$ are orthonormal, that is,
\begin{equation}
\sum_l v_{lk}^*v_{lj}=
{1\over\sqrt{d_jd_k}}\sum_l\tilde c_{jl}\tilde c_{kl}^*=
\delta_{jk}
\end{equation}
(this means, in particular, that the number of rows of $v$ is not smaller 
than the number of columns), and thus by adding columns, $v$ can be 
extended to be a unitary matrix.  By moving the unitary operator $U_j$ 
in Eq.~(\ref{eqtn: TPU}) to the other side of the expression, we obtain
\begin{equation}
T_lP_j=
v_{lj}P_MU_j^\dagger=
v_{lj}U_j^\dagger P_j=
v_{lj}\tilde R_j\;,
\end{equation}
which implies that
\begin{equation}
T_lP_N=
\sum_j T_lP_j=
\sum_j v_{lj}\tilde R_j\;.
\end{equation}
Since the decomposition operators $T_lP_N$ are related to the 
decomposition operators $\tilde R_j$ by a unitary matrix, we can 
conclude, as promised, that ${\cal T}_N$ and ${\cal R}_N$ are the 
same operation.

The upshot is that the part of a reversal operation that actually 
effects the reversal---that is, the restriction of the reversal operation
to the subspace $N$---is uniquely determined.  In what follows this
permits us to make general statements about all reversal operations.

The decomposition used to define ${\cal R}$ in 
Eqs.~(\ref{eqtn: reversal decompositionj}) and 
(\ref{eqtn: reversal decompositionN}) is special.  For any $\rho$ whose 
support is confined to $M$, this decomposition is a canonical decomposition 
for ${\cal R}$ relative to the output state $\rho'={\cal E}(\rho)/\mu^2$ 
of Eq.~(\ref{eqtn: rhoprime}).  This fact is crucial to our later analysis 
of the thermodynamic efficiency of error correction.  To prove it, notice 
that for any $\rho$ whose support is confined to $M$, we have 
$P_j\rho' P_k=\lambda_j\delta_{jk}\rho_j$ and 
$P_{\,\overline N\,}\rho'=\rho' P_{\,\overline N\,}=0$.  It follows 
that the $W$ matrix is diagonal, 
that is,
\begin{equation}
\tilde W_{jk}=
\mbox{tr}(\tilde R_j\rho'\tilde R_k^{\dagger})= 
\mbox{tr}(U_j^\dagger P_j\rho' P_kU_k)=
\lambda_j\delta_{jk}
\end{equation}
and 
\begin{equation}
\tilde W_{j\,\overline N\,} = 
\tilde W_{\,\overline N\,Nj} = 
\tilde W_{\,\overline N\,\overline N\,} = 0 \;.
\end{equation}

The canonical $W$ matrices for ${\cal E}$ and ${\cal R}$ being the same, 
the entropy exchange in the reversal operation is the same as the entropy 
exchange in ${\cal E}$:
\begin{equation}
S_e(\rho',{\cal R})=H(\vec\lambda)=S_e(\rho,{\cal E})\;.
\label{eqtn: reversal entropy exchange}
\end{equation}
In addition, since ${\cal R}(\rho')=\rho$, Eq.~(\ref{eqtn: condition2again})
can be recast as
\begin{equation} 
\label{eqtn: entropy change equals entropy exchange}
S_e(\rho',{\cal R})=S(\rho')-S\bigl({\cal R}(\rho')\bigr)\;.
\end{equation}
This result, that the entropy exchange in the reversal operation is
equal to the entropy reduction, is important for our discussion of 
the thermodynamics of error correction in Sec.~\ref{sect: second law}.  
Since the entropy exchange~(\ref{eqtn: reversal entropy exchange}) is
determined by the restriction of ${\cal R}$ to $N$, 
Eqs.~(\ref{eqtn: reversal entropy exchange}) and 
(\ref{eqtn: entropy change equals entropy exchange}) hold for operation
that reverses ${\cal E}$ on $M$.  We stress, however, that 
Eq.~(\ref{eqtn: entropy change equals entropy exchange}) does not hold 
generally for trace-preserving operations; rather, as 
Eq.~(\ref{eqtn: ArakiLieb}) shows, all that one can say for a general 
trace-preserving operation is that the entropy reduction does not 
exceed the entropy exchange.

We can also make some powerful observations about the reversibility of
entire classes of operations.  Knill and LaFlamme \cite{Knill97a} showed 
that if an operation ${\cal E}$, with decomposition operators $A_k$, is 
reversible on $M$, then any operation ${\cal F}$, whose decomposition 
operators $B_j$ are linear combinations of the $A_k$, is also reversible 
on $M$.  This can be seen immediately from 
Eq.~(\ref{eqtn: algebraic condition}): if 
\begin{equation}
B_j=\sum_k b_{jk} A_k
\end{equation}
(since $b_{jk}$ is not assumed to be a unitary matrix, this is not just
a unitary remixing, which would yield another decomposition of ${\cal E}$ 
instead of a new operation), then 
\begin{equation}
P_MB_k^\dagger B_jP_M=
\sum_{l,m}b_{km}^*b_{jl}P_MA_m^\dagger A_lP_M
=\left(\sum_{l,m}b_{jl}m_{lm}b_{km}^*\right)\!P_M
=n_{jk}P_M\;,
\end{equation}
where the matrix $n=bmb^\dagger$ is manifestly positive.  

This result can be stated compactly in the language of operator subspaces:
the reversibility of an operation ${\cal E}$ on $M$ implies the 
reversibility on $M$ of any operation whose decomposition operators 
span a subspace of the span of the decomposition operators of ${\cal E}$.  
We stress that the decomposition operators $A_k$ can be written as a linear 
combination of the decomposition operators $B_j$ only if the operators $B_j$ 
span the entire operator subspace spanned by operators $A_j$.

We can go further to show that any operation that reverses ${\cal E}$ is 
also a reversal operation for ${\cal F}$.  To do so, notice first that the 
decomposition operators $B_j$ can be written as a linear combination of the 
canonical decomposition operators $\tilde A_k$ of 
Eq.~(\ref{eqtn: restrictedpolar}):
\begin{equation}
B_j=\sum_k\tilde b_{jk}\tilde A_k\;.
\end{equation}
Thus the action of ${\cal F}$ on any density operator $\rho$ whose support 
is confined to $M$ can be written as
\begin{equation}
{\cal F}(\rho)=
\sum_j B_jP_M\rho P_MB_j^\dagger
=\sum_{j,k,l}
\tilde b_{jk}\tilde b_{jl}^*
\tilde A_kP_M\rho P_M\tilde A_l^\dagger
=\sum_{j,k,l}
\sqrt{d_kd_l}\,\tilde b_{jk}\tilde b_{jl}^*
P_kU_k\rho U_l^\dagger P_l
\;,
\end{equation}
The constant value of $\mbox{tr}\bigl({\cal F}(\rho)\bigr)$ is given by
\begin{equation}
\nu^2=\mbox{tr}\bigl({\cal F}(\rho)\bigr)=
\sum_{j,k}d_k|\tilde b_{jk}|^2\;,
\end{equation}
and simple algebra shows that ${\cal R}$ reverses ${\cal F}$:
\begin{equation}
{\cal R}\circ{\cal F}(\rho)=\nu^2 \rho\;.
\end{equation}
Moreover, since only the restriction of ${\cal R}$ to $N$ is involved
in the reversal and any operation that reverses ${\cal E}$ has the
same restriction to $N$, we can conclude that any operation that reverses
${\cal E}$ also reverses ${\cal F}$.  The converse is not true, however,
because the canonical decomposition of ${\cal F}$ might map $M$ to a set 
of orthogonal subspaces whose span is a {\it proper\/} subspace of $N$, 
in which case reversal of ${\cal F}$ would not entail reversal over all 
of $N$.

The reversibility theorem proved in Sec.~\ref{sect: algebraic theorem} 
can be recast in another, very compact algebraic form.  Suppose a quantum 
operation ${\cal E}$, with operator-sum decomposition consisting of 
operators $A_j$, can be reversed on a subspace $M$.  Introduce the 
Hilbert-Schmidt inner product for operators, defined by
\begin{equation}
(N,O) \equiv \mbox{tr}(N^{\dagger} O)\;. 
\end{equation}
This inner product allows us to define an adjoint of a superoperator,
that is, any linear operator on operators.  For example, the adjoint 
of ${\cal E}$ is given by 
\begin{equation}
{\cal E}^{\dagger}(\rho) = \sum_j A_j^{\dagger} \rho A_j\;.
\end{equation}
To see that this is an adjoint with respect to the Hilbert-Schmidt inner 
product, notice that
\begin{equation}
\bigl(N,{\cal E}(O)\bigr)=
\mbox{tr}\!\left(N^{\dagger}\sum_j A_j O A_j^{\dagger}\right) 
=\mbox{tr}\!\left(
\Biggl(\sum_j A_j^{\dagger} N A_j\Biggr)^{\!\dagger}O\right) 
=\bigl({\cal E}^{\dagger}(N),O\bigr)\;,
\end{equation}
which is the required inner-product relation for an adjoint.  The adjoint 
of an operation is generally not an operation, since it can be 
trace-increasing, but it is always a completely positive linear map.

Let ${\cal E}_M$ be the restriction of ${\cal E}$ to the subspace $M$,
as in Eq.~(\ref{eqtn: EsubM}).  Observing that
\begin{equation}
{\cal E}_M^{\dagger}\circ{\cal E}_M(\rho)=\sum_{j,k}
	P_M A_k^{\dagger} A_j P_M \, \rho \, P_M A_j^{\dagger} A_k P_M
\;,
\end{equation}
we see that condition~(\ref{eqtn: algebraic condition}) is equivalent
to the requirement that ${\cal E}_M^\dagger\circ{\cal E}_M$ be a positive 
multiple of the identity operation on $M$.  This requirement can be
written as
\begin{equation}
{\cal E}_M^{\dagger}\circ{\cal E}_M(\rho)=\gamma^2 P_M\rho P_M
\label{eqtn: algebraic condition2}
\;,
\end{equation}
where the positive constant $\gamma^2$ is the trace of $m^2$, that is,
\begin{equation}
\gamma^2=\sum_{j,k}|m_{jk}|^2=\sum_j d_j^2=\mu^4\sum_j\lambda_j^2\;.
\end{equation}
Equation~(\ref{eqtn: algebraic condition2}) is a necessary and sufficient 
condition for reversibility of ${\cal E}$ on $M$.  We note that 
${\cal E}_M^\dagger$ is generally {\em not\/} the reversal operator 
${\cal R}$, for ${\cal E}_M^\dagger$ has decomposition operators
$P_M A_j^\dagger=\sqrt{d_j}P_M U_j^\dagger=\sqrt{d_j}U_j^\dagger P_j$,
whereas the reversal operation has decomposition operators shorn of
the factor $\sqrt{d_j}$.  The omission is necessary so that ${\cal R}$
is a trace-preserving operation.

An equivalent way of writing condition~(\ref{eqtn: algebraic condition2}) 
is to require that ${\cal E}_M$ preserve, up to the constant $\gamma^2$,
the Hilbert-Schmidt inner product of operators defined on $M$; that is,
\begin{equation}
\bigl({\cal E}_M(N),{\cal E}_M(O)\bigr)=
\bigl(N,{\cal E}_M^\dagger\circ{\cal E}_M(O)\bigr)
=\gamma^2 (N,P_M OP_M)
=\gamma^2 (P_M NP_M,P_M OP_M)
\end{equation}
for all operators $N$ and $O$.

\section{Error correction and the Second Law of Thermodynamics}
\label{sect: second law}

Error correction---that is, reversal of an operation---decreases the 
entropy of a quantum system, so it is natural to inquire about the 
thermodynamic efficiency of this process.  In this section we address 
the question of the entropy cost of error correction and show that 
error correction can be regarded as a sort of ``refrigeration,'' 
wherein information about the system, obtained through measurement, is 
used to keep the system cool.  Indeed, the method of operation of an 
error correction scheme is very similar to that of a ``Maxwell demon,''
and the methods of analysis we use are based on those used to resolve
that famous problem.  As a prelude to our analysis, we review and
extend the discussion of the Araki-Lieb inequality found in 
Sec.~\ref{sect: fid-ent}.

\subsection{Useful inequality}
\label{sect: Araki-Lieb}

The Araki-Lieb inequality \cite{Araki70a,Lieb75a,Wehrl78a} states that for 
two systems, $1$ and $2$,
\begin{equation} 
\label{eqtn: Araki-Lieb}
S(1) - S(2) \leq S(12)\;.
\end{equation}
To see this, introduce a third system, $3$, which purifies $12$.  
Subadditivity of the von Neumann entropy and the purity of $123$ imply 
that
\begin{equation}
S(1) = S(23) \leq S(2) + S(3) = S(2) + S(12)\;, 
\label{eqtn: subadd}
\end{equation}
which gives the desired result.  The inequality in 
Eq.~(\ref{eqtn: subadd}) is the statement of subadditivity; equality 
holds if and only if systems $2$ and $3$ are in a product state, that is,
\begin{equation} 
\label{eqtn: Araki-Lieb equality}
\rho^{23} = \rho^2 \otimes \rho^3\;. 
\end{equation}
By interchanging the roles of systems $1$ and $2$ in the above proof, the 
Araki-Lieb inequality can be written more generally as
\begin{equation} 
\label{eqtn: Araki-Lieb 2}
|S(1) - S(2) | \leq S(12)\;.
\end{equation}

Suppose we apply the inequality~(\ref{eqtn: Araki-Lieb 2}) to a 
deterministic quantum operation ${\cal D}$:  
\begin{equation}
|S(\rho^{R'})-S(\rho^{Q'})| \leq S(\rho^{RQ'})\;.
\end{equation}
{}For a deterministic quantum operation, we have that
$S(\rho^{R'})=S(\rho^R)=S(\rho^Q)$; furthermore, it is always true that 
$S(\rho^{RQ'}) = S_e(\rho,{\cal D})$.  Substituting these identities into 
the previous equation gives
\begin{equation}
|S(\rho^Q) - S(\rho^{Q'})| \leq S_e(\rho,{\cal D})\;.
\end{equation}
A special case of this inequality is particularly useful in our entropic 
analysis of error correction:
\begin{equation} 
\label{eqtn: second law}
S_e(\rho,{\cal D})\geq S(\rho^Q) - S(\rho^{Q'})=-\Delta S\;, 
\end{equation}
where $\Delta S \equiv S(\rho^{Q'}) - S(\rho^Q)$ is the change in the
entropy of the system.  From the equality 
condition~(\ref{eqtn: Araki-Lieb equality}), we see that equality holds 
in the preceding equation if and only if 
\begin{equation} 
\label{eqtn: second law equality}
\rho^{QE'} = \rho^{Q'} \otimes \rho^{E'}\;. 
\end{equation}
These equality conditions are crucial to the following analysis of 
thermodynamically efficient error correction.

\subsection{Reversal by a ``Maxwell demon''}

Consider the error-correction ``cycle'' depicted in Fig.~1.  The cycle 
can be decomposed into four stages: 

\begin{enumerate}

\item The system, starting in a state $\rho$, is subjected to a noisy
quantum evolution that takes it to a state $\rho^n$.  We denote the 
change in entropy of the system during this stage by $\Delta S$.  In 
typical scenarios for error correction, we are interested in cases where
$\Delta S \geq 0$, though this is not necessary.

\item  A ``demon'' performs a pure measurement, described by operators 
$\{B_i\}$, on the state $\rho^n$.  The probability that the demon obtains 
result $i$ is 
\begin{equation}
p_i=\mbox{tr}(B_i\rho^n B_i^\dagger)\;,
\end{equation}
and the state of the system conditioned on result $i$ is
\begin{equation}
\rho_i=B_i\rho^n B_i^\dagger/p_i\;.
\end{equation}
All error-correction schemes can be done in such a way that a measurement 
step of this type is included.

\item  The demon ``feeds back'' the result $i$ of the measurement as
a unitary operation $V_i$ that creates a final system state 
\begin{equation}
\rho^c=
V_i\rho_i V_i^\dagger=
V_i B_i\rho^nB_i^\dagger V_i^\dagger/p_i\;,  
\end{equation}
which is the same regardless of which measurement result was obtained. 
In the case of error correction this final state is the ``corrected'' 
state.

\item The cycle is restarted.  In order that this actually be a cycle
and that it be a successful error correction, we must have $\rho^c=\rho$.

\end{enumerate}

\noindent
The second and third stages are the ``error-correction'' stages.  The 
idea of error correction is to restore the original state of the system 
during these stages.  In this section we show that the reduction in 
the system entropy during the error-correction stages comes at the
expense of entropy production in the environment, which is at least
as large as the entropy reduction.

\begin{center}
\begin{picture}(280,300)(0,-100)

\put(0,160){\framebox(70,20){$\rho$}}
\put(200,160){\framebox(70,20){$\rho^n$}}
\put(200,10){\framebox(70,20){``Demon''}}
\put(0,10){\framebox(70,20){$\rho^c$}}
\put(200,-40){\framebox(70,20){Memory: $i$}}
\put(40,-40){\framebox(90,20){Environment}}
 
\put(195,20){\vector(-1,0){120}}
\put(75,170){\vector(1,0){120}}
\put(235,155){\vector(0,-1){120}}
\put(35,35){\vector(0,1){120}}
\put(195,-30){\vector(-1,0){60}}
\put(230,-15){\vector(0,1){20}}
\put(240,5){\vector(0,-1){20}}

{\footnotesize
\put(100,150){\begin{tabular}{c}
         Noisy evolution \\ $(\Delta S \geq 0) $
        \end{tabular} }
\put(158,100){\begin{tabular}{c}
   	Measure $\{ B_i \}$ \\
 	$\rho_i = B_i \rho^n B_i^{\dagger}/p_i$ \\
	$p_i = \mbox{tr}(B_i \rho^n B_i^{\dagger})$
        \end{tabular} }
\put(98,35){\begin{tabular}{c}
        ``Feedback'' result \\
        $\rho_i \rightarrow \rho^c= V_i\rho^i V_i^{\dagger}$
        \end{tabular}}
\put(40,100){\begin{tabular}{c}
	Restart cycle \\ $\rho^c=\rho$
	\end{tabular}}
\put(163,-25){$I_i$}
}

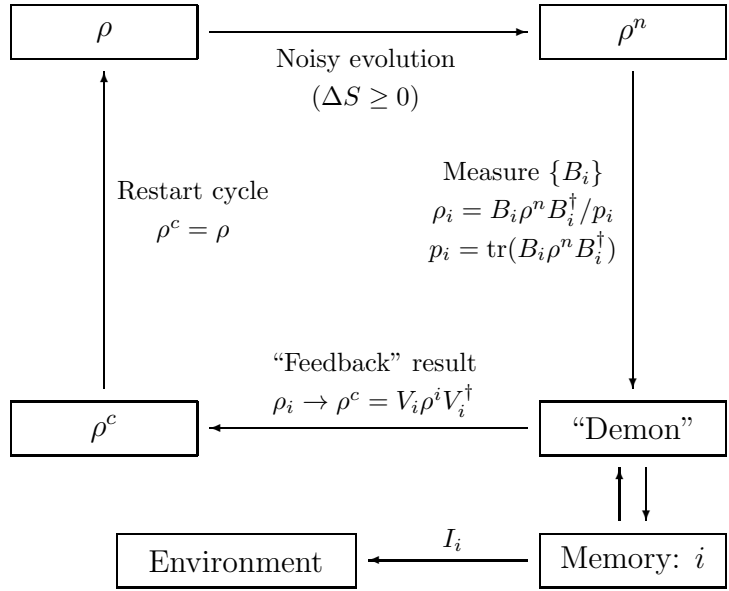
\put(53,-80){Figure~1. Error correction cycle.}
 
\end{picture}
\end{center}

To investigate the balance between the entropy reduction of the system 
and entropy production in the environment, we adopt the ``inside view'' 
of the demon.  After stage~3 the only record of the measurement 
result~$i$ is the record in the demon's memory.  To reset its memory 
for the next cycle, the demon must erase its record of the measurement 
result.  Associated with this erasure is a thermodynamic cost, the 
{\it Landauer erasure cost\/} \cite{Landauer61a,Landauer88a}, which 
corresponds to an entropy increase in the environment.  The erasure cost 
of information is equivalent to the thermodynamic cost of entropy, when 
entropy and information are measured in the same units, conveniently chosen 
to be bits.  Bennett \cite{Bennett82a} used the idea of an erasure cost to 
resolve the paradox of Maxwell demons, and Zurek \cite{Zurek89b} and 
later Caves \cite{Caves90a} showed that a correct entropic accounting 
from the ``inside view'' can be obtained by quantifying the amount of 
information in a measurement record by the algorithmic information
content $I_i$ of the record.  Algorithmic information is the information 
content of the most compressed form of the record, quantified as the
length of the shortest program that can be used to generate the record
on a universal computer.  We show here that the average thermodynamic
cost of the demon's measurement record is at least as great as the 
entropy reduction achieved by error correction.

In a particular error-correction cycle where the demon obtains 
measurement result~$i$, the total thermodynamic cost of the 
error-correction stages is $I_i+\Delta S_c$, where
\begin{equation}
\Delta S_c\equiv S(\rho^c)-S(\rho^n)
\end{equation}
is the change in the system entropy in the error-correction stages.
What is of interest to us is the average thermodynamic cost,
\begin{equation}
\sum_i p_i(I_i+\Delta S_c)=\sum_ip_iI_i +\Delta S_c\;,
\end{equation}
where the average is taken over the probabilities for the measurement
results.  To bound this average thermodynamic cost, we now proceed through
a chain of three inequalities.  

The first inequality is a strict consequence of algorithmic information 
theory: the average algorithmic information of the measurement records 
is not less than the Shannon information for the probabilities $p_i$,
that is,
\begin{equation}
\sum_i p_iI_i\ge H(\vec p\,)=-\sum_ip_i\log p_i\;.
\end{equation}
{}Furthermore, Schack \cite{Schack94b} has shown that any universal computer
can be modified to make a new universal computer that has programs for
all the raw measurement records which are at most one bit longer than 
optimal code words for the measurement records.  On such a modified
universal computer, the average algorithmic information for the measurement 
records is within one bit of the Shannon information $H$.

To obtain the second and third inequalities, notice that the corrected
state $\rho^c$ can be written as
\begin{equation}
\rho^c=
\sum_i p_iV_i\rho_iV_i^\dagger=
\sum_i V_iB_i\rho^n B_i^\dagger V_i^\dagger\equiv
{\cal R}(\rho^n)
\;,
\end{equation}
where ${\cal R}$ is the deterministic reversal operation for the 
error-correction stages.  The operators $V_iB_i$ make up an operator-sum
decomposition for the reversal operation.  The probabilities $p_i$ are 
the diagonal elements of the $W$ matrix for this decomposition,
\begin{equation}
p_i=
\mbox{tr}(B_i\rho^nB_i^\dagger)=
\mbox{tr}(V_iB_i\rho^nB_i^\dagger V_i^\dagger)
\;.
\end{equation}
Thus we have our second inequality from Eq.~(\ref{eqtn: SEminShannon}),
\begin{equation}
H(\vec p\,)\ge S_e(\rho^n,{\cal R})\;.
\end{equation}
Equality holds here if and only if the operators $V_iB_i$ are a canonical
decomposition of ${\cal R}$ with respect to $\rho^n$.  We stress that
different measurements and conditional unitaries at stages 2 and 3 lead 
to the same reversal operation, but they yield quite different amounts 
of Shannon information.

The third inequality is obtained by applying the 
inequality~(\ref{eqtn: second law}) to ${\cal R}$ and $\rho^n$:
\begin{equation} 
\label{eqtn: entropy 2}
S_e(\rho^n,{\cal R}) + \Delta S_c \geq 0\;. 
\end{equation}
As Eq.~(\ref{eqtn: entropy change equals entropy exchange}) shows, 
equality holds here if the operators $V_iB_i$ are a canonical
decomposition of ${\cal R}$.

Stringing together the three inequalities, we see that the total entropy
produced during the error-correction process is greater than or equal to 
zero:
\begin{equation} 
\label{eqtn: second law ok}
\sum_ip_iI_i+\Delta S_c\ge
H(\vec p\,)+\Delta S_c\ge
S_e(\rho^n,{\cal R})+\Delta S_c\ge
0\;.
\end{equation}
Stated another way, this result means that the total entropy change 
around the cycle is at least as great as the initial change in entropy 
$\Delta S$, which is caused by the first stage of the dynamics.  The 
error-correction stage can be regarded as a kind of refrigerator, 
similar to a Maxwell demon, achieving a reduction in system entropy 
at the expense of an increase in the entropy of the environment due to 
the erasure of the demon's measurement record.  

How then does this error-correction demon differ from an ordinary Maxwell 
demon?  An obvious difference is that the error-correction demon doesn't
extract the work that is available in the first step of the cycle as 
the system entropy increases under the noisy quantum evolution.  A subtler,
yet more important difference lies in the ways the two demons return the 
system to a standard state, so that the whole process can be a cycle.  For
the error-correction demon, it is the error-correction steps that reset
the system to a standard state, which is then acted on by the noisy
quantum evolution.  For an ordinary Maxwell demon, the noisy quantum 
evolution restores the system to a standard state, typically thermodynamic 
equilibrium, starting from different input states representing the different 
measurement outcomes.

Can this error correction be done in a thermodynamically efficient manner?
Is there a strategy for error correction that achieves equality in the 
Second Law inequality~(\ref{eqtn: second law ok})?  The answer is yes, 
and we give such a strategy here.  The proof of the Second Law 
inequality~(\ref{eqtn: second law ok}) uses three inequalities, 
$\sum_ip_iI_i\ge H$, $H\geq S_e$, and $S_e \geq -\Delta S$.  To achieve 
thermodynamically efficient error correction, it is necessary and sufficient 
that the equality conditions in these three inequalities be achieved.  

We have already noted that Schack has shown that the first inequality,
$\sum_ip_iI_i\ge H(\vec p\/)$, can be saturated to within one bit by using 
a universal computer that is designed to take advantage of optimal coding 
of the raw measurement records~$i$.  On such a universal computer the 
average amount of space needed to store the programs for the measurement
records---that is, the encoded measurement records---is within one bit of 
the Shannon information $H$.  Moreover, it is possible to reduce this one
bit asymptotically to zero by the use of block coding and reversible 
computation.  The demon stores the results of its measurements using an
optimal code for a source with probabilities $p_i$.  Thus the demon stores
an encoded list of measurement results.   Immediately before performing a 
measurement, the demon decodes the list of measurement results using 
reversible computation.  It performs the measurement, appends the result 
to its list, and then re\"encodes the enlarged list using optimal block 
coding done by reversible computation.  In the asymptotic limit of large 
blocks, the average length of the compressed list of measurement results 
becomes arbitrarily close to $H(\vec p\,)$ per measurement result.

The second inequality, $H(\vec p\,)\ge S_e(\rho^n,{\cal R})$, can be
saturated by letting the measurement operators $B_i$ and conditional 
unitaries $V_i$ be those defined be the those defined by the canonical
decomposition of the reversal operation ${\cal R}$.  It should be noted 
that the optimal method of encoding the measurement records depends on 
the probabilities $p_i$, which in turn are ultimately determined by the 
initial state $\rho$.  Thus the type of encoding needed to efficiently 
store the measurement record generally depends on the initial state 
$\rho$.  For the canonical scheme for error correction, however, the 
probabilities $p_i$ do not depend on the initial state $\rho$. 

The third inequality, 
$S_e({\cal R},\rho^n)\ge -\bigl(S(\rho^c)-S(\rho^n)\bigr)$, 
is satisfied by any error-correction procedure that corrects errors
perfectly.  Indeed, in Sec.~\ref{sect: algebraic discussion} we showed 
that the entropy exchange associated with any reversing operation is 
equal to the entropy reduction achieved by the reversing operation
(see Eq.~(\ref{eqtn: entropy change equals entropy exchange})).
An alternative demonstration that perfect error correction achieves the 
equality $S_e=-\Delta S_c$ begins by noting that at the end of the 
error-correction process $RQ$ must be in a pure state--the initial 
state---and therefore the overall state must be a product 
$\rho^{RQ''}\otimes\rho^{EA''}$ (recall that $E$ is the environment for 
the noise stage, while $A$ is the ancilla for the reversal stage).  Thus 
the condition $\rho^{QA''}=\rho^{Q''}\otimes\rho^{A''}$ certainly holds.  
This is the equality condition~(\ref{eqtn: second law equality}) for the 
Araki-Lieb inequality, applied to the reversal operation.  Hence we have 
$S_e=-\Delta S_c$ for the reversal operation, and we conclude that any 
successful error-correction procedure automatically achieves equality in 
Eq.~(\ref{eqtn: entropy 2}).  It would be interesting to see whether
equality can be achieved in Eq.~(\ref{eqtn: entropy 2}) by error-correction
schemes that do not correct errors perfectly.

\subsection{Discussion}

Zurek \cite{Zurek78a}, Milburn \cite{Milburn96a}, and Lloyd \cite{Lloyd96a} 
have analyzed examples of quantum Maxwell demons, though not in the context 
of error correction.  Lloyd notes that ``creation of new information'' in 
a quantum measurement is an additional source of inefficiency in his scheme, 
which involves measuring $\sigma_z$ for a spin in a static $B$-field applied 
along the $z$ axis, in order to extract energy from it.  If the spin is 
measured in the ``wrong'' basis---for example, if it is initially in a pure
state not an eigenstate of $\sigma_z$---the measurement fails to extract all
the available free energy of the spin, because of the disturbance to the 
system state induced by the measurement.  In the case of error correction,
something similar happens, but it is not disturbance to the system that 
is the source of inefficiency.  Instead, if the ancilla involved in the 
reversal decoheres in the ``wrong'' basis---that is, the measurement 
performed by the demon is not the one defined by the canonical decomposition 
of the reversal operation---then the Landauer erasure cost is greater than 
the efficient minimum $S_e$.  This can be thought of a ``creation of new 
information,'' due to ``disturbance'' of the ancilla, but the change in 
the system state is independent of the basis in which the ancilla decoheres.

Error correction can be accomplished in ways other than that depicted 
in Fig.~1.  The ``inside view'' of the preceding subsection, in which the
demon makes a measurement described by some decomposition of the reversal 
operation, arises when the demon is decohered by an environment, the 
particular measurement being defined by the basis in which the environment
decoheres.  If the demon is isolated from everything except the system and 
is initially in a pure state, then its entropy gain is $S_e = -\Delta S$ 
for the error-correction process.  One can restart the error-correction 
cycle by discarding the demon and bringing up a new demon, the result 
being an increase in the environment's entropy by the demon's entropy 
$S_e$.  This way of performing error correction, which does not involve 
any measurement records, is equivalent to the ``outside view'' of the 
demon's operation.

The ``inside view'' of the demon's operation, we stress again, arises
if the demon's memory is ``decohered'' by interaction with an environment,
the measurement record thus becoming ``classical information.''  In this
case the demon has the entropy $H(\vec p\,)$ of the measurement record, 
not just the entropy $S_e$.  Once this decoherence is taken into account, 
the different decompositions of the reversal operation, corresponding to 
different measurements, constitute operationally different ways of 
reversing things, rather than just different interpretations of the same 
overall interaction.  Keeping in mind the variety of decompositions of 
the reversal operation might lead one to consider a greater variety of 
experimental realizations, some of which may be easier to perform than 
others.  As we emphasize above, a reversal in which the decohered 
measurement results correspond to a canonical decomposition of the 
reversal operation is the reversal method that is most efficient 
thermodynamically.  

\section{Conclusion}
\label{sect: conc}

In this paper we analyze reversible quantum operations, giving both a 
general in\-for\-ma\-tion-theoretic characterization and a general algebraic 
characterization.  Our results help in understanding quantum error 
correction, teleportation, and the reversal of measurements.  By applying 
our two characterizations to a thermodynamic analysis of error correction, 
we show that the reduction in system entropy due to error correction is 
compensated by a corresponding increase in entropy of the rest of the 
world.  Moreover, we show that error-correction schemes that correct 
errors perfectly can be done, in principle, in a thermodynamically 
efficient manner.

\section*{Acknowledgments}

MAN thanks W.~H.~Zurek for thought-provoking discussions concerning
thermodynamics and error correction.  This work was supported in part
by the Office of Naval Research (Grant No.\ N00014-93-1-0116).  MAN,
CMC, and HB thank the Institute for Theoretical Physics for its 
hospitality during the ITP Workshop on Quantum Computers and Quantum
Coherence and for the support of the National Science Foundation 
(Grant No.\ PHY94-07194).  MAN acknowledges financial support from the 
Australian-American Educational Foundation (Fulbright Commission).


\end{document}